\documentclass[10pt,twocolumn,notitlepage,prb,superscriptaddress,longbibliography]{revtex4-2}
\usepackage{here}
\usepackage{amsmath}         
\usepackage{graphicx}
\usepackage{braket}         
\usepackage{lettrine}
\usepackage{xcolor}
\usepackage{hhline}
\usepackage{tabularx}
\usepackage{calc}
\usepackage{soul}

\begin{document}

\title{Promoting $p$-based Hall effects by $p$-$d$-$f$ hybridization in Gd-based dichalcogenides} 

\author{Mahmoud Zeer}
\email{m.zeer@fz-juelich.de}
\affiliation{Peter Gr\"unberg Institute and Institute for Advanced Simulation, Forschungszentrum J\"ulich and JARA, 52425 J\"ulich, Germany}
\affiliation{Department of Physics, RWTH Aachen University, 52056 Aachen, Germany}

\author{Dongwook Go}
\affiliation{Peter Gr\"unberg Institute and Institute for Advanced Simulation, Forschungszentrum J\"ulich and JARA, 52425 J\"ulich, Germany}
\affiliation{Institute of Physics, Johannes Gutenberg-University Mainz, 55099 Mainz, Germany}

\author{Peter Schmitz}
\affiliation{Peter Gr\"unberg Institute and Institute for Advanced Simulation, Forschungszentrum J\"ulich and JARA, 52425 J\"ulich, Germany}
\affiliation{Department of Physics, RWTH Aachen University, 52056 Aachen, Germany}

\author{Tom G. Saunderson}
\affiliation{Institute of Physics, Johannes Gutenberg-University Mainz, 55099 Mainz, Germany}

\author{Hao Wang}

\affiliation{Peter Gr\"unberg Institute and Institute for Advanced Simulation, Forschungszentrum J\"ulich and JARA, 52425 J\"ulich, Germany}

\author{Jamal Ghabboun}
\affiliation{Department of Physics, Bethlehem University, Bethlehem, Palestine}

\author{Stefan Bl\"ugel}
\affiliation{Peter Gr\"unberg Institute and Institute for Advanced Simulation, Forschungszentrum J\"ulich and JARA, 52425 J\"ulich, Germany}

\author{Wulf Wulfhekel}
\affiliation{Physikalisches Institut, Karlsruhe Institute of Technology, 76131 Karlsruhe, Germany}

\author{Yuriy Mokrousov}
\email{y.mokrousov@fz-juelich.de}
\affiliation{Peter Gr\"unberg Institute and Institute for Advanced Simulation, Forschungszentrum J\"ulich and JARA, 52425 J\"ulich, Germany}
\affiliation{Institute of Physics, Johannes Gutenberg-University Mainz, 55099 Mainz, Germany}

\begin{abstract}

%We investigate the potential benefits of replacing transition metal with rare-earth elements in two-dimensional (2D) structures of transition metal dichalcogenides (TMDs). To this end,
We conduct a first-principles study of Hall effects in rare-earth dichalcogenides, focusing on monolayers of the H-phase EuX$_2$ and GdX$_2$, where X = S, Se, and Te. Our predictions reveal that all  EuX$_2$ and GdX$_2$ systems exhibit high magnetic moments and wide bandgaps.
We observe that while in case of EuX$_2$ the $p$ and $f$ states hybridize directly below the Fermi energy, the absence of $f$ and $d$ states of Gd at the Fermi energy results in $p$-like spin-polarized electronic structure of GdX$_2$, which mediates $p$-based magnetotransport. 
%Furthermore, we consider the effect of the Coulomb repulsion (U) parameter and observe that a large value of U leads to strong hybridization of the $p$-$f$-$d$ orbitals in the case of EuX$_2$, while a large value of U causes the minority/majority density of states (DOS) of the $p$-state to shift down/up when the $f$ state of Gd is shifted in energy. 
Notably, these systems display significant anomalous, spin, and orbital Hall conductivities.
We find that in GdX$_2$ the strength of correlations controls the relative position of $p$, $d$ and $f$-states and their hybridization which has a crucial impact on $p$-state polarization and the anomalous Hall effect, but not the spin and orbital Hall effect.
Moreover, we find that the application of strain can significantly modify the electronic structure of the monolayers, resulting in quantized charge, spin and orbital  transport in GdTe$_2$ via a strain-mediated orbital inversion mechanism taking place at the Fermi energy. %Furthermore, strain induces quantized values of AHC and quantized plateaus of SHC and OHC in close proximity to the Fermi energy. 
Our findings suggest that rare-earth dichalcogenides hold promise as a platform for topological spintronics and orbitronics.

\end{abstract}

\date{\today}                 
\maketitle

\section{Introduction}

In recent years, motivated by their potential applications in spintronics and quantum computing, significant advances have been made in exploring magnetism of two-dimensional (2D) materials. %Notably, the discovery of ferromagnetism in monolayer CrI$_3$ \cite{Huang2017} has opened up a new avenue for exploring 2D magnetism. 
Since the discovery of ferromagnetism in monolayer CrI$_3$ \cite{Huang2017}, a plethora of 2D magnetic materials has been discovered, including monolayer Fe$_3$GeTe$_2$ \cite{Wang2020}, and monolayer FePS$_3$ \cite{olsen2021}. Among novel trends,  the utilization of 4$f$ electrons from rare-earth atoms in 2D materials, such as for example rare-earth dichalcogenides, offers distinct advantages for advanced magnetic storage devices and spintronics applications. The localized 4$f$ electrons mediate high magnetic moments, which contribute to  robust magnetocrystalline anisotropies when combined with the strong spin-orbit coupling (SOC) and unique crystal field in 2D geometry~\cite{wangGd2020,liu2021,sheng2022,you2021}.

In 2D materials, in addition to the anomalous and spin Hall effects, (AHE and SHE), %The Spin Hall effect (SHE) is a phenomenon in which a transverse spin current is generated by a longitudinal charge current in a non-magnetic (NM) material with strong spin–orbit coupling (SOC) 
\cite{kato2004, slawinska2019, safeer2019, guo2005}, the orbital Hall 
effect (OHE), i.e. the generation of trasnverse flow of orbital angular momentum, started attracting considerable attention. 
%Similarly, the Orbital Hall effect (OHE) generates an orbital current, which is a flow of orbital angular momentum, and exhibits distinct features compared to the SHE. 
Unlike the SHE, the OHE is often dominated by a contribution from non-relativistic momentum-space orbital textures and can be found in multi-orbital systems irrespective of the strength of spin-orbit coupling \cite{go2018}. Additionally, theoretical calculations have shown that the OHE can be by far dominant over SHE in a variety of materials \cite{go2021}, providing a more efficient way to manipulate and detect quantum states in devices. The investigation of OHE has covered various materials, including 3$d$-5$d$ transition metals \cite{tanaka2008, Jo2018}, graphene, and two-dimensional transition metal dichalcogenides (TMDs) \cite{canonico2020, costa2022, cysne2021,Bhowal2020}.
%. In TMDs, the OHE is caused by the non-zero Berry curvature of the spin-split bands, which is generated by the spin-valley coupling \cite{Bhowal2020}.
The OHE can be also utilized for magnetization control in  spin-orbit torque  devices relying on an injection of an orbital current, which often originates from light elements and their surfaces~\cite{Go2020, ramaswamy2018, go2020orbital, smaili2021,ding2020harnessing,ding2022observation,Saunderson2022}. %In SOT devices, the spin-polarized current generated by the OHE can switch the magnetization of a ferromagnetic layer, thereby changing the magnetoresistance of different spin channels and providing potential applications in spintronics devices \cite{Go2020, ramaswamy2018, go2020orbital, smaili2021}.
Overall, the OHE emerges as a promising phenomenon for spintronics applications.
%due to its distinct features compared to the SHE. It can be tuned by external electric fields and can be used to generate spin-polarized currents in both magnetic and non-magnetic layers.

Recently, we theoretically demonstrated that the monolayer EuS$_2$ exhibits intriguing electronic and magnetic properties, and hosts large anomalous, spin and orbital Hall effects~\cite{PhysRevMaterials.6.074004}. 
%We found that the electronic structure of EuS$_2$ is highly sensitive to the Coulomb repulsion parameter $U$, which governs the hybridization between Eu-$f$ and S-$p$ states and influences the system's transport properties. 
%Specifically, we observed that the hybridization between Eu-$f$ and S-$p$ states is strongly dependent on the value of $U$, resulting in significant changes in the electronic structure of the system. Moreover, we 
%revealing that the monolayer EuS$_2$ displays remarkable values of anomalous, spin, and orbital Hall conductivities. 
These findings motivated us to further investigate the electronic structure and transport behavior of two-dimensional 4$f$ rare-earth dichalcogenides, (REDs). Here, we focus in detail on the physics of
%the structural, electronic, magnetic, and transport properties of 
H-phase monolayers of  EuSe$_2$, EuTe$_2$, GdS$_2$, GdSe$_2$, and GdTe$_2$, using first-principles calculations. Our goal is to explore the potential of these materials as sources of pronounced charge, spin, and orbital currents, taking into account the correlation effects.

 Our results reveal that all considered REDs  exhibit large magnetic moments and band gaps for a wide range of correlation strength.
 We find that while the properties of Eu-based REDs are governed by direct interaction of $p$- and $f$-states below the Fermi energy, for the case of REDs which are based on Gd, it is purely the $p$-states, which are well-separated from the $f$- and $d$-states but are spin-polarized by them, that determine the electronic structure and transport properties. We demonstrate that prominent charge, spin and orbital response takes place in considered materials upon applying an electric field, uncover the tunability of transport and electronic properties by correlations, and promote strain as a powerful tool in driving topological Hall response of GdTe$_2$, which can be turned into a Chern and quantum spin Hall insulator by a parent orbital Hall insulator phase at the Fermi energy. This shows that complex $p$-$d$-$f$-hybridization taking place in REDs can result in purely $p$-based physics of topological transport in time-reversal broken $f$-based 2D phases.  
 %Interestingly, we find that the band gap increases with increasing $U$ for the Eu-based systems, while it changes only slightly for the Gd-based systems. 
 %Moreover, the magnetic moment shows minimal changes with varying $U$ for all systems. 
 %Additionally, we observe that spin-orbit coupling (SOC) has a significant impact on the electronic structure of the Eu/Gd-based systems, leading to a large spin-splitting of the bands near the Fermi level.
 %Furthermore, we find that the transport properties of the Eu/Gd-based systems are strongly influenced by the correlation strength $U$.
%Additionally, we discuss the potential of strain as a tool to manipulate the charge, spin, and orbital currents in GdTe$_2$ monolayers. 
%Our findings demonstrate that compressive strain plays a crucial role in shifting and generating substantial peaks of charge, spin, and orbital currents at the Fermi energy position.
Our findings suggest that  REDs hold promise for spintronics and magnetotransport applications. %However, we emphasize that further research is needed to fully comprehend their potential in integrating into existing semiconductor technologies. 
%This study underscores the importance of investigating the properties of REDs monolayer systems to gain a comprehensive understanding of their suitability for spintronics applications.  

Our manuscript is structured as follows: Section II provides detailed information on the computational methodology used in this study. Section III presents the results of our calculations and analysis, accompanied by a discussion of the findings. Finally, in Section IV, we provide a concise conclusion summarizing our research. For the sake of convenience, in the following, we will refer to the monolayers of EuS$_2$, EuSe$_2$, and EuTe$_2$ as EuX$_2$ (X = S, Se, Te) and the monolayers of GdS$_2$, GdSe$_2$ and GdTe$_2$ as GdX$_2$ (X= S, Se, Te).

\section{Computational Details}

We conducted our first-principles calculations using the density functional theory (DFT) code FLEUR~\cite{wortmann2023fleur}, which implements the full-potential linearized augmented plane wave method~\cite{Wimmer1981}. The Perdew-Burke-Ernzerhof approximation~\cite{Perdew1996} was used to account for exchange and correlation effects. The monolayer systems were relaxed to obtain their optimal structures, both with and without the inclusion of the Coulomb interaction strength $U$, as shown in Table~\ref{table:Lattice}. The SOC effect was considered using the second-variation scheme in all calculations.
We performed self-consistent calculations  using a 16$\times$16$\times$1 Monkhorst-Pack grid in the first Brillouin zone. We found that all compounds favor  ferromagnetic ground state.

To construct maximally localized Wannier functions (MLWFs) from the Bloch wave functions obtained from the self-consistent DFT calculation, we utilized the Wannier90 package
~\cite{pizzi2020wannier90}. For the EuX$_2$ systems, the MLWFs were constructed from the $f$ and $d$ orbitals of the Eu atom, and the $p$ orbitals of the X atom. For the GdX$_2$ systems, the MLWFs were constructed from the $d$ orbitals of Gd and the $p$ orbitals of the X atom, as the $f$ state of Gd is located far away from the Fermi energy (see below). The maximum frozen window was set to be 2 eV higher than the Fermi energy for each system.
A mesh of 8$\times$8$\times$1 k-points was employed to obtain 36 and 22 MLWFs for Eu and Gd systems, respectively.
The matrix elements of spin and orbital angular momentum operators were first evaluated in the Bloch basis and then transformed into the MLWF basis.

To account for the effect of electronic correlations, we utilized the GGA+$U$ method within the self-consistent DFT iteration. The Hubbard $U$ correction was employed to address the strong on-site Coulomb repulsion between the electrons in the 4$f$-shell of Eu and Gd. We selected the on-site Coulomb interaction strength $U$ to be 0, 2.5, and 6.7 eV~\cite{Shick1999,Kurz2002}, and the intra-atomic exchange interaction strength $J$ to be 0.7 eV~\cite{Shick1999,Kurz2002,carbone2022}. It is worth noting that a more accurate description of our systems could be achieved by using the widely accepted value of 6.7 eV for Eu/Gd. Nevertheless, we also performed calculations with a smaller value of $U$=2.5 eV to gain a general understanding of the interplay between bonding and correlations in these systems.
 
\section {Results and discussion}
 \subsection{Electronic and magnetic properties of EuX$_2$/GdX{$_2$} }

 \subsubsection{Electronic structure of EuX{$_2$} }
 
 \begin{figure*}[ht!]
    \includegraphics[angle=0, width=0.95\textwidth]{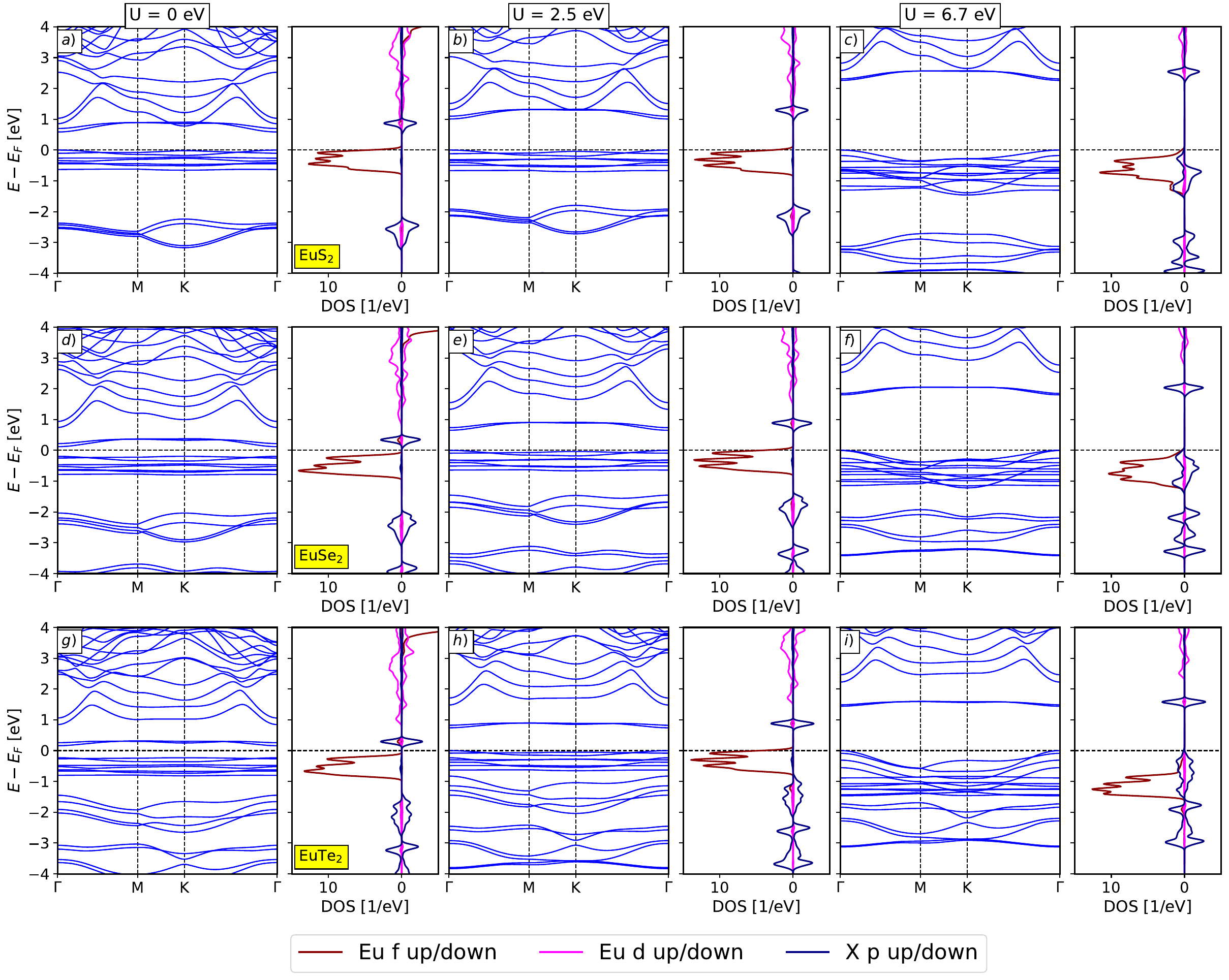}
    \caption{
    %\textcolor{red}{The DOS need better colors. RGB ! Not easily visible / catchy ! Have to cut band structures to the necessary region. AND: compute f/p-projection of DOS normed to 1. plot total DOS seperately in black !}.
    Electronic structure of the H-phase $\text{EuX}_2$ monolayers as a function of Coulomb repulsion strength $U$ of 0 (left column), 2.5 (middle column) and 6.7\,eV (right column), as shown in (a)-(c), (d)-(f), and (g)-(i) for $\text{EuS}_2$, $\text{EuSe}_2$, and $\text{EuTe}_2$, respectively. Left and right parts of each figure correspond to the  band structure and spin-resolved density of states (DOS). %correspond to the majority and minority spins, respectively. 
    The DOS of Eu-$f$, Eu-$d$, and X-$p$ states is shown in dark red, pink, and blue, respectively.
%The hybridization of Eu-$f$ states with X-$p$ states is further enhanced by the presence of Eu-$f$ states located at the Fermi level with $p$-states of X atoms. This hybridization is also evident in the density of states (DOS).
}
    \label{fig:EuX2_bandDOS}
\end{figure*}
 
The computed structural and magnetic properties of EuX$_2$ (X = S, Se, Te)  are presented in Table~\ref{table:Lattice}. We find that the local Eu magnetic moment in these systems is around  6.95\,$\mu_\mathrm{B}$, being quite independent of the correlation strength and chemical composition. The local magnetic moment of chalcogen atoms was found to be small. When going from S to Te, we notice an expansion of the lattice, and an increase in the distance between the planes of  Eu and X atoms, Table~\ref{table:Lattice}, which can be explained by an increasing spread of the chalcogen $p$-states with increasing the atomic number.

The computed band structures of EuX$_2$ (X = S, Se, Te) are shown in Fig.~\ref{fig:EuX2_bandDOS} (a)-(i), exhibiting a band gap at the Fermi energy for all systems. For $U=0$\,eV the valence band maximum originates from the majority 4$f^{\uparrow}$ states of the Eu atom, while the conduction band bottom consists of the $p$ states of chalcogenide atoms. Increasing the value of $U$ from 0 to 6.7\,eV leads to an increase in the band gap without a radical change in dispersion. For instance, the band gap increased from 0.59, 0.31 and 0.39 to 2.27, 1.81 and 1.45 eV for EuS$_2$, EuSe$_2$ and EuTe$_2$ respectively, while the valence band maximum of EuTe$_2$ acquires predominantly Te-$p$ character. Additionally, increasing $U$ brings the majority of 4$f^{\uparrow}$-states down in energy which results in strong direct hybridization between the 4$f^{\uparrow}$ Eu and the X $p$ states. This hybridization has a strong impact on the transport properties, as discussed further. Overall, the basic ingredients of the electronic structure of EuX$_2$ have been analyzed on the case of EuS$_2$ in our previous work, Ref.~\onlinecite{PhysRevMaterials.6.074004}.

When comparing the three compounds with each other, several key differences can be observed. First of all, the relative distance between the valence $f$- and lower $p$-states decreases with increasing the atomic number when going from S to Te via Se. And while with increasing $U$ the $p$- and $f$-states are ultimately brought into the same region of energy, in case of EuTe$_2$ the $p$-states ``overshoot" the $f$-states, get promoted to the Fermi energy, and push Eu $f$-bands down in energy significantly. This can be explained by the overall reduction in repulsive interaction between the lowest conduction and highest valence $p$-bands. In turn, this results in the reduction of the band gap when going from EuS$_2$ to EuTe$_2$, see Table~\ref{table:Lattice}. Secondly, the small exchange splitting of  chalcogen $p$-states, visible in the DOS, increases with increasing the atomic number. Since the $p$-magnetism in our systems arises due to $p$-$d$-$f$ hybridization, larger spread of $p$-orbitals of Te as compared to that of S $p$-bands leads to stronger hybridization and correspondingly larger spin splitting of $p$-states. The variation in spin splitting has direct impact on the transport properties of $p$-states, as discussed further. 

%For EuTe$_2$, the top $p$-states of Te are not touching the majority of $f$ $-$state of Eu, thus not influencing any transport effect of AHC, SHC, and OHC. In the case of EuSe$_2$, however, the hybridization of the 4$f$ state of Eu and the $p$ state of Se atoms leads to the formation of a pronounced peak of OHC at the Fermi energy. The common view of the band structures of each material at $U$=0 show distinct features, including $p$ bands just above the Fermi level and below the large $f$ bands. Upon increasing $U$, the band structure of each material evolves in a similar manner, with the $p$ orbitals shifting upwards through the Fermi level and hybridizing with the $f$ orbitals at $U$=6.7. Moreover, the $f$-state of Eu goes down in energy upon changing the  chalcogenide atom.

%The effects of modifying the chemical composition, particularly through the increase of the atomic number of X in EuX$_2$ systems, are conspicuously observed in the relative energetic positioning of the $p$-state bands of the X atoms. With increasing atomic number, the global gaps between the $f$-$p$ states near the Fermi energy decrease, as shown in Fig.~\ref{fig:EuX2_bandDOS} (c), (f) and (i). On the other hand, the magnetic properties of EuX$_2$ (X = S, Se, Te) have also been studied and the results are presented in Table \ref{table:Lattice}. The magnetic moment of these systems is found to be 6.95\,$\mu_\mathrm{B}$, which is relatively constant and independent of the correlation strength and chemical compositions.

\begin{table*}[t]

\caption{Relaxed atomic positions and magnetic properties of Eu/GdX{$_2$} compounds. Shown are the values of the lattice constant (a),~distance between $M$ and $X$ atomic planes $\Delta{(M-X)}$,  magnetic moments of  M and X atoms, and the values of the bandgap. The value of $U$ is taken to be 6.7\,eV.}

\begin{tabular}[t]{lccccc}

\hline
\hline
&Lattice constant&$\Delta{(M-X)}$&Magnetic moment (M)&Magnetic moment (X)& band gap \\
&[\AA]&[\AA]&[$\mu_B$] & [$\mu_B$] & [eV] \\

\hline
   
\hline
EuS$_2$&4.616&1.103&6.82&$-$0.010&0.59\\
EuS$_2$+$U$&4.744&1.080&6.95&$-$0.028& 2.27\\
EuSe$_2$&4.824&1.209&6.81&$-$0.005&0.31\\
EuSe$_2$+$U$&4.907&1.237&6.95&$-$0.024&1.81\\
EuTe$_2$&5.036&1.440&6.83&$-$0.009&0.39\\
EuTe$_2$+$U$& 5.177&1.417&6.91&$-$0.026&1.45\\
GdS$_2$&4.138&1.388&6.86&$-$0.35&0.82\\
GdS$_2$+$U$&4.192&1.400&7.05&$+$0.32&0.74\\
GdSe$_2$&4.301&1.492&6.83&$-$0.37&0.62\\
GdSe$_2$+$U$&4.265&1.521&7.05&$+$0.33&0.53\\
GdTe$_2$&4.571&1.654&6.84&$-$0.32&0.37\\
GdTe$_2$+$U$&4.550&1.693&7.11&$+$0.27&0.34\\
\hline
\hline
 \label{table:Lattice}
\end{tabular}
\end{table*}%

\subsubsection{Electronic structure of GdX{$_2$} }

  \begin{figure*}[ht!]
    \includegraphics[angle=0, width=0.95\textwidth]{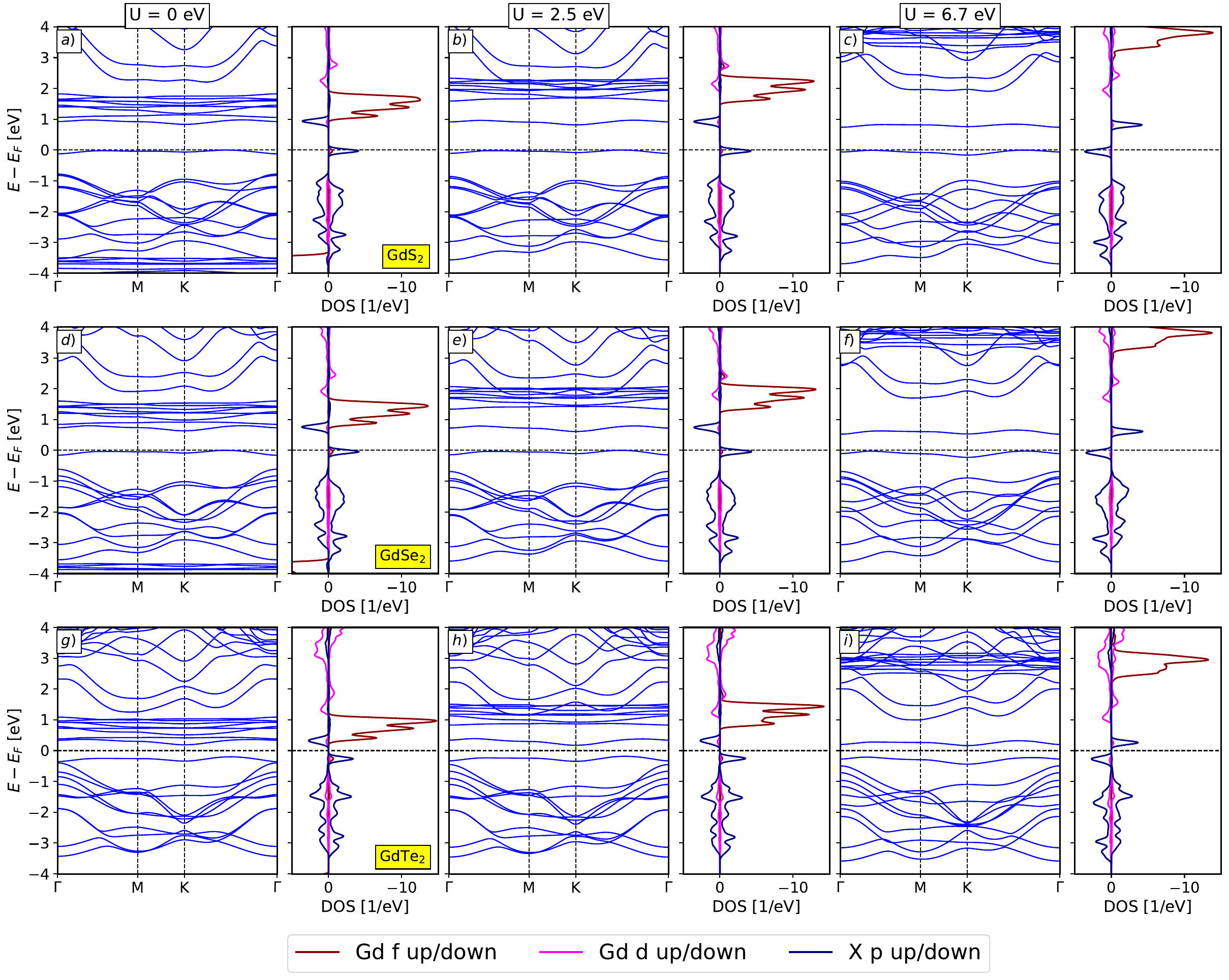}
    \caption{
    Electronic structure of the H-phase $\text{GdX}_2$ monolayers as a function of Coulomb repulsion strength $U$ of 0 (left column), 2.5 (middle column) and 6.7\,eV (right column), as shown in (a)-(c), (d)-(f), and (g)-(i) for $\text{GdS}_2$, $\text{GdSe}_2$, and $\text{GdTe}_2$, respectively. Left and right parts of each figure correspond to the  band structure and spin-resolved DOS. %correspond to the majority and minority spins, respectively. 
    The DOS of Gd-$f$, Gd-$d$, and X-$p$ states is shown in dark red, pink, and blue, respectively.
    %Electronic structure of H-phase $\text{GdX}_2$ monolayer as a function of Coulomb repulsion strength $U$.
    %In (a)-(c), (d)-(f) and (g)-(i) the band structures and the corresponding evolution of the spin-resolved density of states for $U=0$, $2.5$= and $6.7\ \mathrm{eV}$ of $\text{GdS}_2$, $\text{GdSe}_2$ and $\text{GdTe}_2$ are shown respectively. The left and right parts of the all figures beside the band structure correspond to the majority and minority spins, respectively. while, the density of states of Gd-$f$, Gd-$d$, X-$S$ states is shown with dark red, green and blue, respectively. 
    Upon increasing $U$ the minority/majority Gd-$f$ states shift up/down in energy and the minority/majority X $p$-states shift down/up. The $p$-$d$-$f$ hybridization between the Gd and  X atoms below the Fermi energy transforms the separate group of $p$-states into a transport-active group of states. }
    \label{fig:GdX2_bandDOS}
\end{figure*}
 
We next discuss the electronic structure of GdX$_2$ compounds, shown in Fig.~\ref{fig:GdX2_bandDOS}. We find that the maximum of the valence band of GdX$_2$ is located along the $\Gamma$ $-$K path, while the minimum of the conduction band is positioned at the K point, resulting in an indirect, small band gap formed by the occupied and unoccupied  $p_z$-states of X atom with opposite spin. Thus, all GdX$_2$ compounds exhibit a typical bipolar magnetic semiconducting feature, where the highest valence band and the lowest conduction band have an opposite spin orientation~\cite{li2012}. Such a feature is believed to offer significant advantages for the development  of spintronics nanodevices~\cite{cheng2018}.
%and the advancement of quantum information processing \cite{cheng2018}.
We also investigate the effect of on-site Coulomb interaction $U$ on the band gap of GdX$_2$. At $U=0$\,eV, the magnitude of the indirect band gaps of GdX$_2$ systems is 0.82, 0.62, and 0.37\,eV for GdS$_2$, GdSe$_2$, and GdTe$_2$, respectively. We observe that with increasing $U$ the band gap changes only slightly, see Table~\ref{table:Lattice}. %Furthermore, all GdX$_2$ compounds exhibit a typical bipolar magnetic semiconducting feature, where the highest valence band and the lowest conduction band have completely opposite spin orientations \cite{li2012}. Consequently, these bipolar magnetic semiconductors offer significant advantages for the development of next-generation spintronic nano devices and the advancement of quantum information processing \cite{cheng2018}. 

In contrast to Eu-based REDs considered above, the Gd majority (fully occupied) and minority (empty) $f$-states are positioned quite far away from the Fermi energy, and thus the properties around the Fermi energy are determined by the X $p$-states. The hybridization between these $p$-states  and the $d,f$-states  of Gd is evident from the DOS in the energy range from $-$4 eV to $-$1 eV. This hybridization is particularly strong near the Fermi level,
%, where the $p$-orbitals of X atoms are strongly coupled with the $d$-orbitals of Gd. 
and it is largely responsible for the observed  transport properties, as we shall see below. Key for magnetotransport properties of GdX$_2$ is the induced spin-splitting of the $p$-states. Around the Fermi energy, a sizeable spin splitting and spin-dependent occupation of the $p_z$ states is largely responsible for an overall much larger moment of the X atom in GdX$_2$ as compared to the case of EuX$_2$, see Table~\ref{table:Lattice}.

%of the GdX$_2$ systems where the magnetic coupling is ascribed to 5 $d$ of Gd and $p$ of X atoms mediated 4$f$–4$f$ exchange interactions. 
Let us analyze in detail the evolution of the electronic structure and $p$-magnetism as we increase the effect of correlations. By increasing the value of $U$ from 0 to $2.5$ and then further to $6.7$ eV, we observe a clear trend in the hybridization between the Gd-$f,d$ states and X-$p$ states, which is reflected in their energetic position.
%of the minority/majority DOS of the $p$ state.
Specifically, we find that the majority(${\uparrow}$)/minority(${\downarrow}$) DOS of the Gd-$f$ states moves down/up as we increase the $U$, and the $p$-states follow the same trend, Fig.~\ref{fig:GdX2_bandDOS}(c,f,i). For smaller values of $U$, the $p$-states are spin-polarized with a moment opposite to that of Gd atom. This is the consequence of the $p$-states being positioned in the gap between strongly exchange-split $f$-states, which is best understood by looking at the $p_z$-states positioned right at the Fermi energy: starting from initially spin-degenerate $p$-states here, we realize that as a result of hybridization of majority $p^{\uparrow}$-states with the low $f^{\uparrow}$-states the former will move up in energy, while the minority $p^{\downarrow}$-states will move down in energy as a result of hybridization with high-lying $f^{\downarrow}$-states. This result in the occupation of the  $p^{\downarrow}$-states at the Fermi energy, which also explains the opposite $p$-moment of X atoms.

Moreover, upon increasing $U$, the $f^{\downarrow}$-sates move higher in energy, and the energetic position of the $p$ states starts to be effected by the unoccupied exchange-split Gd $d$-states, lying higher than the $p$-states. This results in an increased $pd$-hybridization as visible also from the partial DOS in Fig.~\ref{fig:GdX2_bandDOS}. Here, as a result of this hybridization, $d^{\uparrow}$-states push  $p^{\uparrow}$-states much lower in energy than the $d^{\downarrow}$-states push down  $p^{\downarrow}$-states, owing to the fact that $d^{\downarrow}$ are situated further from the $p$-states than $d^{\uparrow}$-states. As a result, the polarization of $p$-states follows that of the Gd $d$-states. This promotes the occupation of $p^{\uparrow}$ states and the corresponding X-moment switches its sign as $U$ is increased. This behavior is reflected by deeper lying 
 occupied $p$-states, as can be seen in the DOS. Our calculations thus underline a crucial influence of the subtle details of $pdf$-hybridization for the $p$-magnetism of GdX$_2$ compounds.

 \subsection {Transport properties of EuX$_2$ and GdX$_2$}

 \begin{figure*}[t!]
    \includegraphics[angle=0, width=0.98\textwidth]{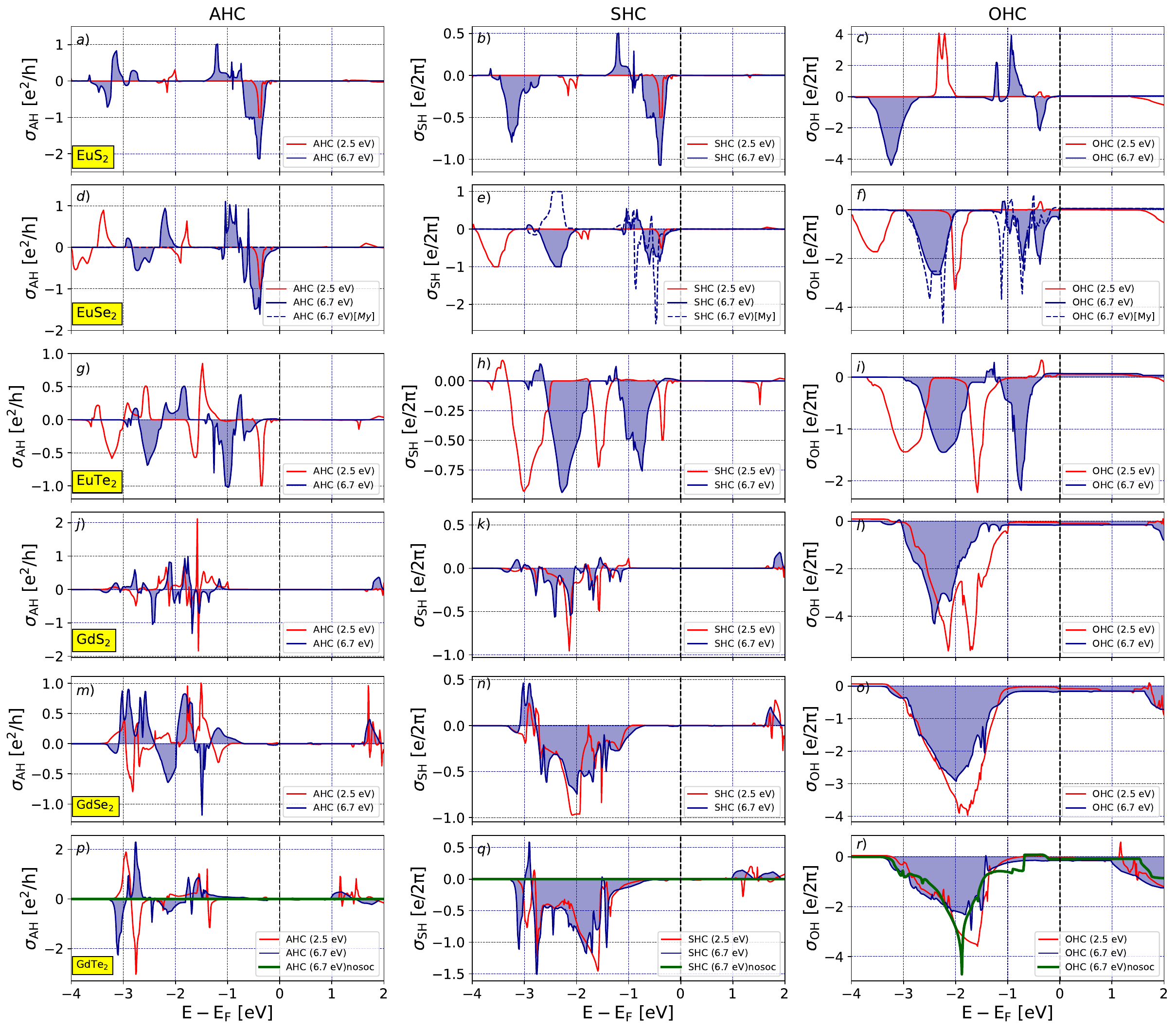}
    \caption{Transport properties of EuX$_2$ and GdX$_2$ as a function of band filling.  Each row corresponds to a different system, with the first column representing the anomalous Hall conductivity (AHC, $\sigma_{\rm AH}$), the second column the spin Hall conductivity (SHC, $\sigma_{\rm SH}$), and the third column the orbital Hall conductivity (OHC, $\sigma_{\rm OH}$),
    %The AHC $\sigma_{\rm AH}$, SHC $\sigma_{\rm SH}$, and $\sigma_{\rm OH}$ OHC of EuX$_2$ and GdX$_2$ are 
    computed for two values of $U$: 2.5\,eV (red dashed line) and 6.7\,eV (blue solid line). For the case of GeTe$_2$ thick  green lines show the values computed without spin-orbit coupling (for $U=6.7$\,eV). For EuSe$_2$ blue dashed line shows the values for the case of in-plane magnetization along $y$ (for $U=6.7$\,eV).
    % The influence of Eu-X hybridization at $U=6.7$\,eV is clearly visible in the plot: while in the case of weak hybridization at $U=2.5$\,eV the AHE and SHE in the system is suppressed, in the case of Eu-X hybridization it shifts or changes the transport more dramatically than Gd-X hybridization.
    }
    \label{fig:Transport}
\end{figure*}

  We investigate the transport properties of EuX$_2$ and GdX$_2$ compounds by computing the intrinsic Berry curvature contribution to the AHC, SHC and OHC  from the MLWF description. To do so, we construct a MLWF Hamiltonian and use the Wannier interpolation technique to calculate the $xy$-component of AHC, SHC, and OHC as a Brillouin zone integral on a 300$\times$300 mesh of $k$ points.
 In case of AHC we use the following expression: 
 \begin{equation}
\sigma_{\rm{AH}}=-\frac{e^{2}}{\hbar}\sum_n\int_\text{BZ} \frac{d^2\boldsymbol{k}}{(2\pi)^2}f_{n\boldsymbol{k}}\,\Omega_{n\boldsymbol{k}},
\label{eq:anomloushall}
\end{equation}
with $f_{n\boldsymbol{k}}$ as the Fermi-Dirac distribution function, and the Berry curvature $\Omega_{n\mathbf{k}}$ of a state $n$ at point $\boldsymbol{k}$ given by
\begin{equation}
\Omega_{n\boldsymbol{k}}=2\hbar^{2} \sum _{n\neq m}
\mathrm{Im}
\left[
\frac{\left<u_{n\boldsymbol{k}} \left|v_{x}\right|u_{m\boldsymbol{k}}\right>\left< u_{m\boldsymbol{k}}\left|v _{y}\right|u_{n\boldsymbol{k}}\right>}{(E_{n\boldsymbol{k}}-E_{m\boldsymbol{k}}+i\eta )^2}
\right],
 \end{equation}
where $E_{n\boldsymbol{k}}$ is the energy of a Bloch state with lattice periodic part of Bloch wave function given by $u_{n\boldsymbol{k}}$,  
and $v_{i}$ is the $i$'th Cartesian component of the velocity operator. For improving the convergence, we set $\eta=25\ \mathrm{meV}$.

The SHC and OHC are assessed using the Kubo expression for the case of transverse in-plane current with out-of-plane spin/orbital polarization:
 \begin{equation}
\sigma_{\text{OH/SH}} =
\frac{e}{\hbar}
\sum_{n}\int \frac{d^{2}\boldsymbol{k}}{(2\pi)^{{2}}}f_{n_{\boldsymbol{k}}}\,\Omega_{n\boldsymbol{k}}^{J_z},
\end{equation}
where so-called spin (orbital) Berry curvature reads:
\begin{equation}
\Omega_{n\boldsymbol{k}}^{J_z}=
2\hbar^{2}\sum_{m\neq n} 
{\rm Im}
\left[
\frac{
\left \langle 
u_{n\boldsymbol{k}}
| j_{y}^{J_z}| 
u_{m\boldsymbol{k}}
\right\rangle
\left \langle  
u_{m\boldsymbol{k}} 
|v_x| 
u_{m\boldsymbol{k}} 
\right \rangle
}
{
(E_{n\boldsymbol{k}} - E_{m\boldsymbol{k}}+i\eta )^2
}
\right]
,
\end{equation}
with $j_{y}^{J_z}$ as the spin ($J_z=S_z$, the $z$ component of the spin operator) or orbital ($J_z=L_z$, the $z$ component of the local angular momentum operator) current operator defined as $j_y^{J_z} = (v_y J_z + J_z v_y)/2$.
 
%We reported the results of EuX$_2$ and GdX$_2$ in separate next sections in the case of $U$=2.5 eV and 6.7 eV. 

We present the results of our conductivity calculations in Fig.~\ref{fig:Transport}. We divide the figure into three columns: the first column presents AHC values, the second $-$ SHC, and the third column $-$ the values of the OHC. Within each panel, the band filling dependence is shown for two values of $U$: 2.5\,eV and 6.7\,eV. Overall, all compounds exhibit significant values of the conductitivies, and below we present a comparative analysis of the transport characteristics in relation to electronic structure features  for each material. 

\subsubsection{Transport properties of EuX$_2$}

%The analysis of Fig~\ref{fig:Transport} for EuX$_2$ shows that among the three compounds, EuS$_2$ exhibits the highest values for AHC, SHC, and OHC. 
%Furthermore, EuSe$_2$ shows slightly higher values compared to EuTe$_2$. 
For the transport properties of EuX$_2$ the $p$-$f$-hybridization plays a crucial role, as discussed in our previous work~\cite{PhysRevMaterials.6.074004}. In both EuS$_2$ and EuSe$_2$ the increase in $U$ drives the shift of the occupied $p$-states up in energy thus fostering a direct overlap and hybridization between $p$- and $f$-states right below the Fermi energy. This leads to a drastic enhancement of AHC, SHC and OHE in the region of energies above $-$1.5\,eV upon changing $U$ from 2.5 to 6.7\,eV, Fig.~\ref{fig:Transport}(a-f). Below $-$2\,eV the transport in EuS$_2$ and EuSe$_2$ is dominated by the spin-polarized $p$-states of X atoms, which carry significant AHC, SHC and OHC, and which move up in energy with increasing $U$. 

In case of EuSe$_2$ for $U=6.7$\,eV, Fig.~\ref{fig:Transport}(d-f), at the energy of $-$2.3\,eV the $p_x,p_y$-states form a global gap which can be characterized with a zero Chern number (i.e. vanishing AHC), but which exhibits quantized values of the SHC and OHC. This manifests the emergence of a combined  time-reversal broken spin Hall insulating and  orbital Hall insulating phase. Remarkably, the character of this topological gap can be tuned by bringing the magnetization to lie in the plane of the monolayer: in this case the AHC vanishes from symmetry, with the quantized value of the OHC remaining intact, but with the spin Chern number changing its sign, Fig.~\ref{fig:Transport}(d-f). At the same time the overall band filling dependence of SHC and OHC in EuSe$_2$ changes drastically upon changing the magnetization direction. Notably, among the compounds studied, EuSe$_2$ is the only system that exhibits a magnetization-direction tunable sharp peak in OHC precisely at the Fermi energy.

The case of EuTe$_2$, Fig.~\ref{fig:Transport}(g-i), stands out quite distinctly when compared to EuS(Se)$_2$.
Indeed, upon changing the magnitude of $U$ from 2.5 to 6.7\,eV, the change in the band-filling dependence of the conductivities, at first sight, can be understood as a simple shift in energy.
Upon inspection of the correlation between the peaks in the conducitivities with evolution of the DOS discussed above, we realize that, as in the case of EuS(Se)$_2$, at smaller value of $U$ the overall behavior can be decomposed into to separate regimes: below $-1$\,eV, determined by $p$-states, and between $-1$\,eV and the Fermi energy, governed by the $f$-states. However, when the $U$ in increased to 6.7\,eV, the highest occupied $p$-states overshoot the $f$-states in EuTe$_2$, with the latter moving down in energy into the region of $[-1.5;-1.0]$\,eV, keeping the direct hybridization with the $p$-states to a minimum.

As a result of this, qualitatively, the contribution of the $p$-states simply shifts in energy with increasing $U$, and the peak due to $f$-states shifts down in energy,  getting modified in the process by the interaction with a small fraction of $p$-states remaining below $-1$\,eV. For example, the $f$-peak in the AHC at $-0.3$\,eV [$U=2.5$\,eV, Fig.~\ref{fig:Transport}(g)] moves to $-1.25$\,eV, the corresponding negative peak in the SHC [$U=2.5$\,eV, Fig.~\ref{fig:Transport}(h)] is recognizable as a small positive peak at $-1.25$\,eV for $U=6.7$\,eV, and a small positive OHC $f$-peak at $-0.3$\,eV [$U=2.5$\,eV, Fig.~\ref{fig:Transport}(i)] moves to $-1.25$\,eV while keeping sign and magnitude. Overall, the $p$-contributions to transport remain by far dominant in magnitude.

  \subsubsection{Transport properties of GdX$_2$}
 
 The transport properties of GdX$_2$, presented in Fig.~\ref{fig:Transport}(j-r), are distinctly different from EuX$_2$, but qualitatively similar among each other.
 Let us first look at the AHC, Fig.~\ref{fig:Transport}(j,m,p). A prominent AHC signal is observed in a wide range below the Fermi energy, and it can be associated with the $p$-states of X, considering that the occupied majority $f$-states are positioned very deep in energy, as discussed above. Not only for the AHC, but also for the SHC and OHC this range of ``Hall-active" energies corresponds to the width of the $p$-band, with an exclusion of the $p_z$-states, spin-split across the Fermi energy, which do not contribute to the Hall effects. The fine structure of the AHC within the $p$-band differs among the members of the GdX$_2$ family, however, the influence of $U$ on the AHC can be understood from the changes in the spin-resolved character of the $p$-states discussed above. Indeed, increasing the $U$ from 2.5 to 6.7\,eV interchanges the position of $p^{\uparrow}$ and $p^{\downarrow}$ states, which results in the reversal of the $p$ spin-polarization. In a simple picture, this should result in a change in the sign of the AHC at a given energy $-$ the qualitative validity of this prediction can be indeed checked in Fig.~\ref{fig:Transport}(j,m,p).

The effect of $U$ is much weaker on the SHC, Fig.~\ref{fig:Transport}(k,n,q). Unlike for the case of EuX$_2$, the increase in $U$ does not result in significant changes of energy positions of the $p$-states below the Fermi energy, while the slight rearrangements of the states within the $p$-band influence the fine structure of the SHC, with an overall SHC magnitude increasing and the distribution becoming more uniform with increasing the atomic number of the X atom. This may be attributed to two factors: (i) a slighly increasing $p$-band width and correspondingly stronger spin-orbit splittings of the states with increasing the atomic number, which drive a stronger SHC; (ii) the two separate groups of $p$-states which separately provide contributions to the SHC and OHC at different energies of EuX$_2$ are coming together within the same energy region for GdX$_2$, contributing by a wide plateau of the conductivities. Unlike the AHC, the overall sign of the SHC stays predominantly negative as a function of band filling and $U$. 

This is even more pronounced in the case of OHC for all GdX$_2$ materials: within the considered range of energies we predict only a negative sign of the OHC, Fig.~\ref{fig:Transport}(l,o,r). In comparison to the SHC behavior, and even more so when compared to the AHC, the shape of the OHC dependence on band filling seems almost universal. One reason for such universality is certainly the fact that the position of the $p$-states and their orbital composition remains frozen when varying the X element $-$ the effect is further enhanced by the fact that the primary OHC roots in the electronic structure of the $p$-states already without spin-orbit interaction, which demolishes the effect of atomic-number dependent band splittings via SOC, on the orbital Hall effect. To demonstrate this clearly, for the case of GdTe$_2$, we compute the conductivities without spin-orbit, finding that while the AHC and SHC expectedly turn to zero, the OHC remains influenced very little by such a drastic change in the electronic structure. Of course, $X$-dependent variation in the OHC can be also observed $-$ note for example the double peak structure of the OHC for GdS$_2$ and its evolution with $U$ which can be attributed to the energy shift of the anticrossing at $K$ from $-2$\,eV [Fig.~\ref{fig:GdX2_bandDOS}(b)] to $-2.5$\,eV [Fig.~\ref{fig:GdX2_bandDOS}(c)] $-$   but this occurs on a much smaller scale than that of the characteristic peak OHC magnitude of the order of 2 to 4\,$e^2/h$.  
%Notably, despite the OHC almost disappearing in the band gap region, a small value is still present in the case of EuTe$_2$. Additionally, our results reveal quantized AHC and SHC effects.
%Thirdly, the magnitude of OHC decreases as the chemical composition of GdX$_2$ is altered, as observed for EuX$_2$. For example, the OHC magnitude of GdS$_2$ is greater than that of GdSe$2$ and GdTe$2$. Conversely, the magnitude of SHC exhibits the opposite trend, increasing with the change in X atom. We observe a quantized SHC of $\sigma{\rm SH} \approx -1,\frac{e}{2\pi}$ in GdSe$_2$ monolayer systems at a Fermi energy (E$_F$) of $-$2 eV, which shifts to $-$2.4 eV with varying $U$ from 2.5 eV to 6.7 eV. 
%Furthermore, we investigate the systems EuX$_2$ and GdX$_2$ without spin-orbit coupling (SOC), which results in zero AHC and SHC and larger values of OHC compared to the case with SOC. We show only the results of GdTe$_2$ as shown in Fig.~\ref{fig:Transport} (p), (q) and (r).
%Furthermore, we investigate the EuX$_2$ and GdX$_2$ monolayer systems in the absence of SOC, leading to the complete suppression of both the AHC and SHC. However, the OHC exhibits significantly higher values compared to the case with SOC. We present the results only for GdTe$_2$, as shown in  Figure~\ref{fig:Transport} (p), (q), and (r).

\begin{figure}[t!]
    \includegraphics[angle=0, width=0.45\textwidth]{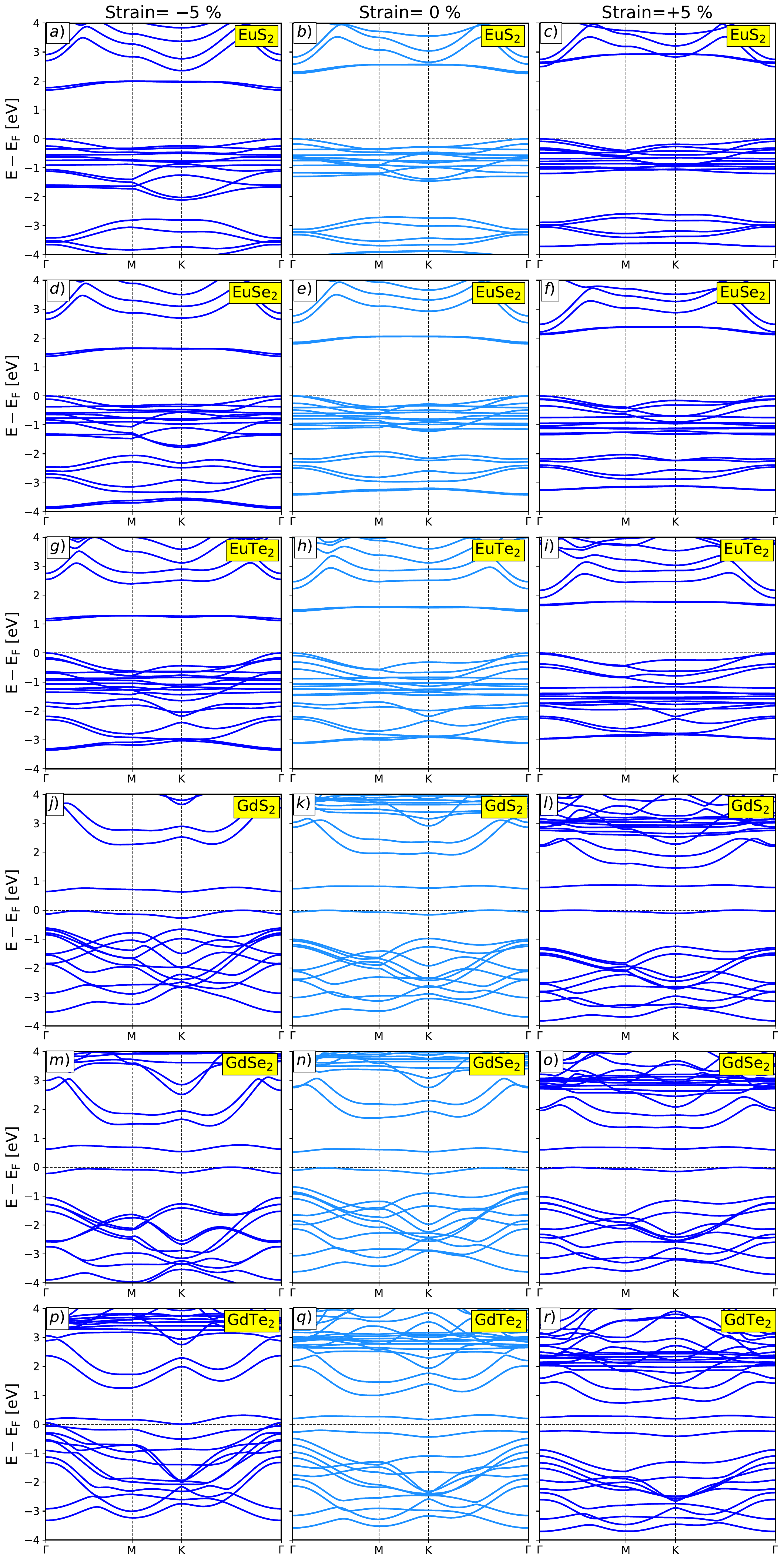}
%    {Fig4.pdf}
    \caption{Evolution of the band structure of EuX$_2$ (a-i) and GdX$_2$ (j-r)  under $+5$\%  tensile strain (right  panel) and $-5$\% compressive strain (left  panel) in comparison to the case in equilibrium, shown in the middle in faded blue.}
    \label{fig:strain}
\end{figure}

\subsection{Tuning electronic structure by strain}
  
\subsubsection{Electronic structure of strained EuX$_2$ and GdX$_2$}
Next, in our calculations we apply a bi-axial tensile and compressive strain of $\pm$5\%, relax the atomic positions and compute the electronic structure of EuX$_2$ and GdX$_2$,  presenting the evolution of band structures with strain along  K-$\Gamma$  in Fig.~\ref{fig:strain}. % biaxial ? => do map of xy strain !
We observe that despite the influence of strain on the exact values of the magnetic anisotropy energy,  the magnetic moment for all systems remains basically unaffected in size and orientation.
%pointing to a strong interlocking between the occupied orbital texture and spin density.
%with  $\theta_{min}=X=\text{const}$ and 
In EuX$_2$, Fig.~\ref{fig:strain}(a-i), strain exerts a strong influence on the relative energetic positions of the $p$- and $f$-states: with increasing tensile strain the $p$-states generally become less dispersive and move higher in energy, while at the same time the occupied $f$-states move down in energy. This reduces the hybridization and overlap of lowest valence $p$-states with the $f$-bands, has a strong impact on the dominant orbital character of the states around the Fermi energy as the lattice is expanded, and increases the bandgap from 2.27, 1.81 and 1.45 eV at equilibrium to 2.52, 2.13 and 1.65 eV at $+$5\% tensile strain for EuS$_2$, EuSe$_2$ and EuTe$_2$, respectively. Such behavior might be desirable for optoelectronic applications where spin- and orbitally-resolved details of coupling of optical pulse to electronic states may be tuned by strain.

%conductions does alter the superposition of Eu-$f$ and X-$p$ as it gives them an opposite penalty, with Eu-$f$ shifting down / up in energy especially when bound to a heavier chalcogenide and X-$p$ shifting up/down upon $+/- 5\%$ deformation, shown as the thick metallic valence region vs. the degenerate flat CB + dispersing split VB in Fig.~\ref{fig:GdX2_bandDOS} (a)-(i). Thus, as expansion brings the VB X-$p$ states to the Fermi level, it changes the character of conducting and excitable electrons, decreases the overlap with f-states and increases the bandgap, from 2.27, 1.81 and 1.45 eV at equilibrium to 2.52, 2.13 and 1.65 eV at $+$ 5\% tensile strain for EuS$_2$, EuSe$_2$ and EuTe$_2$. 

In case of GdX$_2$, Fig.~\ref{fig:strain}(j-r), the key features of the electronic structure remain intact in the considered strain range: for the majority of the cases, the central group of half-filled exchange-split $p_z$-states positioned right at the Fermi energy remains well-separated from the unoccupied $d$-states and occupied  $p_{x,y}$-bands. This is despite the fact that with increasing tensile strain the $d$-states approach the $p_z$-group, while the $p_{x,y}$-bands come close to the Fermi energy under compressive strain. This means that the profoundly metallic behavior with tunable orbital character at the Fermi energy may be achieved GdS$_2$ and GdSe$_2$ under very large strain. It is only for the case of GdTe$_2$ that at $-$5\% strain the $p_{x,y}$-states reach the Fermi energy which leads to an intricate hybridization with the $p_z$-states and significant modifications in the electronic structure. We consider this case in detail below.    

%Surprisingly, the Gd-based structures have much smaller spectral response to lateral tensile and compressive strain for the occupied $p$-states, remaining an indirect bandgap semiconductor, which suggests that bonding is performed by deeper, unaffected $pxy$ and magnetic expression is mostly carried by $pz$ orbitals. This is supported by the occupation of only one spin-polarized, near-flat X-$p$ band near the Fermi energy. For the heaviest compound, GdTe$_{2}$, a $-5\%$ compression lets the valence $pxy$ and conduction $pz$ bands touch in a complicated mechanism illustrated in fig. \ref{fig:Transport_GdTe2_transition}, resulting in the conversion of the system to an itinerant magnet.

%\begin{figure}[ht!]
%    \centering
%    \includegraphics[angle=0, width=0.45\textwidth]{transition_condpol.png}
%    \caption{Hall conductivities in GdTe2 upon strain and associated band inversion mechanism: While the peaks in AHC, SHC and OHC can be shifted, the plateaus only form between certain bands, here colored by Orbital polarization. Arrows mark the flat band inversions. On the side, the corresponding polarization in the BZ around the Gamma point is depicted in a sketch for each of the 3 involved bands. }
%    \label{fig:Transport_GdTe2_transition}
%\end{figure}

\subsubsection{Strained GdTe$_2$}
\begin{figure}[]
        \includegraphics[angle=270, width=0.44\textwidth]{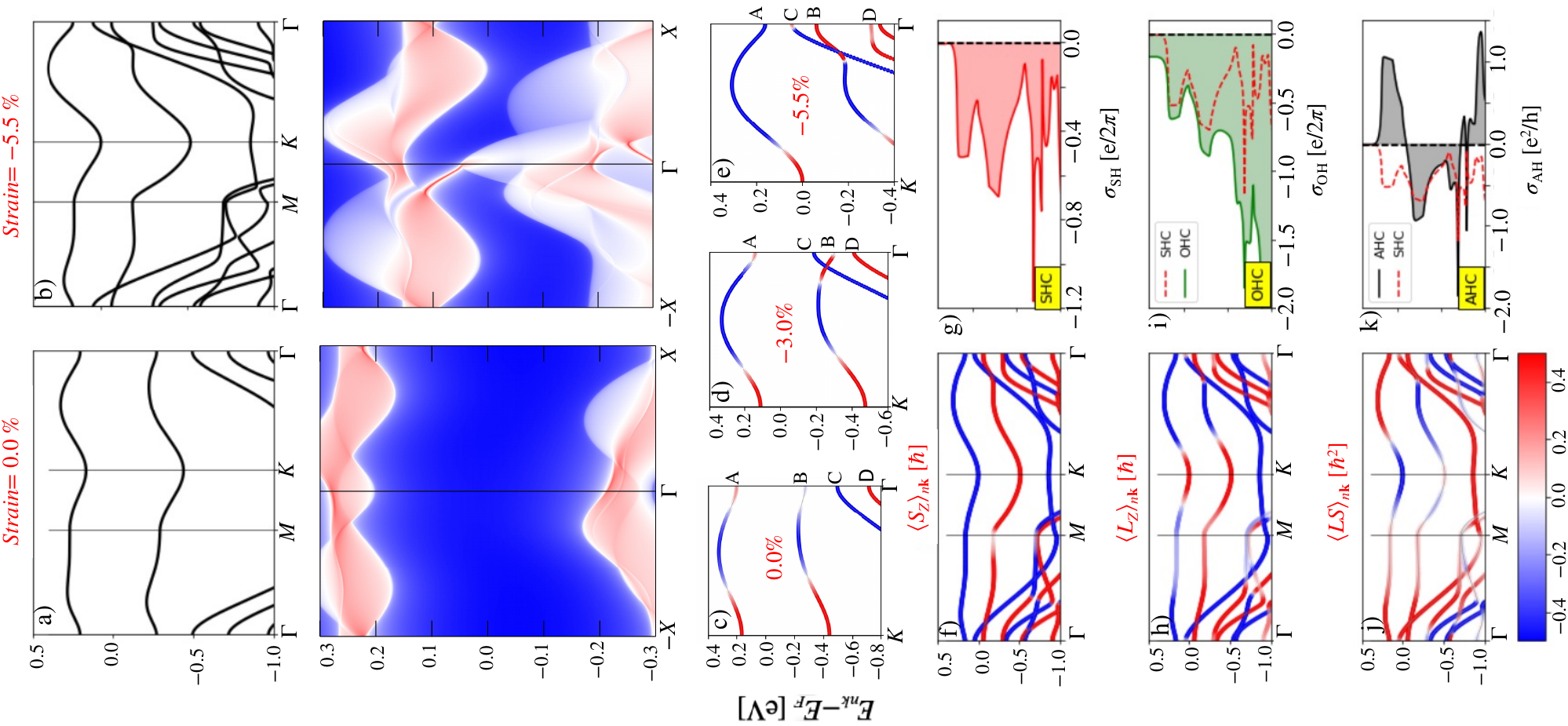}
        \caption{Electronic structure and transport properties of GdTe$_2$. Band structure is shown for the monolayer and for the ribbon without (a) and with (b) compressive strain of $-$5.5\%.  The color range from blue to red represents increasing value of local DOS at one edge of the ribbon. Topological edge states are clearly visible in the strained case around the $\Gamma$ point. (c-e) Shown is the evolution of the bands around the Fermi energy with strain. A, B, C and D mark the  bands of interest, colored by the value of the orbital polarization, mediating band inversion and quantum anomalous Hall effect. 
        %Next to each band a sketch of corresponding distribution of orbital polarization  around the $\Gamma$-point is shown. 
        (f)-(g) show the spin expectation values and spin Hall conductivity,  (e)-(f) show the orbital expectation value and orbital Hall conductivity, and (g)-(h) show the spin-orbit polarization with anomalous Hall conductivity, as a function of energy. In (i) and (k) dashed line stands for the distribution of the SHC.
        %Panels (i), (j), and (k) represent the charge, spin, and orbital Berry curvature, respectively.
        }
        \label{fig:edgestate}
\end{figure}

The closing of the bandgap in GdTe$_{2}$ upon $-5\%$ strain with simultaneous presence of magnetization opens the possibility to implement a quantum anomalous Hall phase after a band inversion which opens a global band gap at the Fermi energy for a slightly larger compression of $-5.5\%$, see Fig.~\ref{fig:edgestate}(a,b). 
Our results show that compressive strain shifts the occupied states towards the Fermi energy and induces changes in their spectral shape. At $-5.5\%$ strain we observe quantized and plateau-like Hall conductivities with values of $\sigma{\rm ^{AH}}\approx 1\frac{e^2}{h}$, $\sigma{\rm ^{SH}}\approx -1\frac{2e^2}{h}$ and $\sigma{\rm ^{OH}}\approx 1\frac{2e^2}{h}$, within the global band gap at the Fermi level, see Fig.~\ref{fig:edgestate}(g,i,k).
%, this region becomes a strong candidate for edge transport controlled by the $p_z$ vs. $pxy$ band competition around $\Gamma$. 
The quantization of the Hall conductance comes with the associated edge states upon strain with a single chiral edge mode visible at $\Gamma$, while the unstrained spectrum stays gapped.
%,Indeed, while the edge spectrum is gapped at equilibrium, we observe a single massless and chiral edge state at the $\Gamma$ point for $-5.5 \%$ compression.
The computation of edge states for a zigzag terminated monolayer was conducted using the Green function method as implemented in WannierTools~\cite{sancho1985highly,wu2018wanniertools}.   

%To confirm the presence of such phase, we compare for $0 \%$ and $-5.5 \%$ the AHC, SHC and OHC, see (Fig.~\ref{fig:Transport_GdTe2_transition}),  the associated Berry Curvature throughout the BZ at the Fermi level (see Fig.~\ref{fig:edgestate} i-k), as well as the edge spectrum (Fig.~\ref{fig:edgestate} (a),(b) and Fig.~\ref{fig:Transport_GdTe2_transition} ). 
%Methodologically we first compute the MLWFs\cite{Niu2020,Niu2019} and then either perform a response calculation with OrbiTrans (LinkDongwook), or a semi-infinite slab diagonalization within WannierTools. It is important to recall that Hall-conductivities correspond to currents at the edge and not in the bulk of the insulating material. 
%Our results show that compressive strain shifts the conducting states towards EFermi and induces changes in their spectral shape. As for $-5.5\%$ strain we observe quantized and plateau-like Hall conductivities with values of $\sigma{\rm ^{AH}}\approx 1\frac{e^2}{h}$, $\sigma{\rm ^{SH}}\approx -1\frac{2e^2}{h}$ and $\sigma{\rm ^{OH}}\approx 1\frac{2e^2}{h}$, within the global band gap at the Fermi level, this region becomes a strong candidate for edge transport controlled by the $p_z$ vs. $pxy$ band competition around $\Gamma$. Indeed, while the edge spectrum is gapped at equilibrium, we observe a single massless and chiral edge state at the $\Gamma$ point for $-5.5 \%$ compression.

To scrutinize the nature of Hall transport properties, we take a closer look at the behavior and correlation between the three types of Hall conductivity, shown in relation to the state-dependent out-of-plane spin polarization, $\Braket{S_z}_{n\mathbf{k}}$, orbital out-of-plane polarization $\Braket{L_z}_{n\mathbf{k}}$ and spin-orbit polarization, defined as an average of the spin-orbit part of the Kohn-Sham Hamiltonian, $\Braket{\mathbf{LS}}_{n\mathbf{k}}$, shown respectively in Fig.~\ref{fig:edgestate}(f), (h) and (j). From inspecting the spin-polarization of the states, Fig.~\ref{fig:edgestate}(f), we observe that despite the presence of spin-orbit coupling, the states remain purely spin-polarized almost everywhere, which leads us to a conclusion that the effect of spin-flip part of spin-orbit interaction on Hall effects is negligible, and that it is primarily the orbital intermixing and hybridization which drive the Hall response. Indeed, the distribution of orbital polarization and its interchange among the bands appears to be much more complex, Fig.~\ref{fig:edgestate}(h).
%Since we found multiple quantized conductivities, it is a-priori unclear whether Magnetization, Spin-Orbit coupling or Orbital band inversion are responsible for the observed edge state. 
%Thus, we investigate the transition in more detail by plotting the Spin- and Orbital polarization as well as the strength of Spin-Orbit coupling for bands around $\Gamma$. 
%As the bands are spin-polarized without color-change due to spin-mixing, while the orbital polarization displays such, we can infer that the Berry curvature originates from orbital hybridization. 
Correspondingly, following the mechanism of orbital inversion, present already without SOC, a strong orbital Hall effect in the vicinity of the Fermi energy of strained GdTe$_2$ arises, Fig.~\ref{fig:edgestate}(i). In turn, upon including spin-orbit into consideration, this gives rise to the SHC, shown in Fig.~\ref{fig:edgestate}(g). 

According to an early model-based theory~\cite{kontani2009}, the SHC and OHC at a given energy  are related to each other by the averaged spin-orbit polarization at that energy.
%pet
%as $SHC=LS\cdot OHC$
%pet
And while it is clear that such a relation cannot be true in general, it seems to qualitatively hold in considered case. Indeed, from looking at the OHC and SHC plotted together in Fig.~\ref{fig:edgestate}(i), we conclude that while the general features in both quantities agree very well, a drop in SHC magnitude at lower energies can be attributed to a suppression of the spin-orbit polarization by the bands with negative  $\Braket{\mathbf{LS}}_{n\mathbf{k}}$ below $-$0.3\,eV, Fig.~\ref{fig:edgestate}(j), with this quantity being predominantly positive above this energy. In turn, given the suppression of spin-flip spin-orbit, the two-channel model of spin Hall and anomalous Hall effect, implying that the two types of conductivities are related by the spin-polarization of the states, %$AHC=\frac{SHC}{m_S} = \frac{OHC}{m_L}$, 
also clearly works in our case.
When plotted together in Fig.~\ref{fig:edgestate}(j), the AHC and SHC exhibit a clear correlation in magnitude and features, and the changes in sign of the AHC can be understood from the changes in the spin-polarization of the states (shown in Fig.~\ref{fig:edgestate}(f)), which carry the corresponding Berry curvatures.

%we can deduce the conductivities from OHC, as
%\begin{align*}
%SHC=LS\cdot OHC \\
%AHC=\frac{SHC}{m_S} = \frac{OHC}{m_L}  
%\end{align*}
%In the topological phase at $-5.5\%$ strain the LS term is positive both in the valence- and conduction band. Consequently, since the integrated OHC is -0.5, the corresponding SHC is negative as well, and the AHC results from dividing this by $+/-\frac{1}{2}$ in the VB/CB around $\Gamma$, leading to a twice as large AHC of -2 in the VB and +2 in the CB. 

Our Berry curvature analysis in $k$-space (not shown)
%pet: Why not ?
indicates that the origin of the topological properties around the Fermi energy comes from interaction among the bands around the $\Gamma$-point, and thus we look closer at the band evolution taking place around $\Gamma$ in response to strain, Fig.~\ref{fig:edgestate}(c-e). 
At equilibrium, Fig.~\ref{fig:edgestate}(c), we see two weakly dispersing around the Fermi energy bands, A and B, with two lower bands C and D separated from band B by a couple of hundreds of meV. Given the predominantly spin-conserving nature of spin-orbit interaction at the Fermi energy, among these four participating bands, the pairs of bands of the same spin, A and C (majority), and B and D (minority), are allowed to hybridize, and the degree of their hybridization increases upon compression of the lattice, as it is already visible for the case of $-$3\% strain (Fig.~\ref{fig:edgestate}(d)), as the bands C and D move up in energy. At the latter strain the hybridization is not strong enough to cause an orbital inversion at $\Gamma$, and the Chern number in the gap remains zero. However, as the bands are brought closer to each other for larger strain, the orbital exchange among bands takes place, Fig.~\ref{fig:edgestate}(e): the reversal in sign of orbital polarization at $\Gamma$ among the bands A and C is clearly visible, while the orbital polarization of band B is increased at the expense of band D loosing it. This orbital inversion process not only explains the rise and quantization of Hall effects in the gap at the Fermi energy, but also can be named responsible for enhanced conductivities in the energy range in between the band edges of B and D bands at the $\Gamma$-point around $-$0.2\,eV. The fact that the AHC would be quantized in that gap given the absence of other ``contaminating" metallic bands is supported by the observation of a distinct edge state around $\Gamma$ between $-0.3$ and $-0.2$
\,eV, Fig.~\ref{fig:edgestate}(b).

\section{Discussion} 

With our work, we have firmly established that if produced (see discussion in Ref.~\cite{PhysRevMaterials.6.074004} complemented by an on-going experimental effort~\cite{ekimov2023}), the  Eu- and Gd-based REDs in the H-phase are not only strongly magnetic but can also give rise to remarkable Hall properties, driven by the details of $p$-$d$-$f$-hybridization and tunable by the chemical composition, degree of correlations and strain. Moreover, we have identified two candidates, EuSe$_2$ and GdTe$_2$, where the electronic structure may result in observable quantization phenomena and corresponding exotic edge physics. In GdTe$_2$, we discover that by inducing the hybridization between the $p_{x,y}$ and $p_z$ states at the Fermi energy via strain, orbital inversion and orbital exchange process can be realized, which in turn leads to the quantization of anomalous, spin and orbital Hall conductivities. The corresponding transport measurements could be directly used to confirm the topological nature of the gap in strained  GdTe$_2$ (e.g. after deposition on a suitable insulating substrate).

On the other hand, in the case of EuSe$_2$, which 
also exhibits a topologically non-trivial gap well below the Fermi energy, spectroscopy techniques such as ARPES~\cite{mlynczak2022fe} could be employed to confirm the presence of the gap and its impact on the edge state physics~\cite{chen2022spanning}. The fact that the AHE in the gap turns to zero irrespective of the magnetization direction, but the quantized values of the SHC and OHC are susceptible to magnetization changes, is unusual. In the past, the variation of the Chern number in 2D magnets within the gap has been attributed to the formation of so-called mixed Weyl points in the combined space of magnetization direction and the $k$-vector, which also manifest in enhanced spin-orbit torques~\cite{niu2019mixed,hanke2017mixed}. Our observations point to an exciting possibility of extending the ``mixed" viewpoint to the case of spin and orbital channels intertwined with $k$-space geometry in 2D magnets, and exploring the physics of Berry-phase driven spin-orbit and orbital torques, either hidden~\cite{Saunderson2022}, or promoted by breaking of inversion symmetry in e.g. Janus-type of monolayers~\cite{liu2022janus,lu2017janus,hernandez2022janus,smaili2021,chen2022intrinsic}. Further investigations in this direction may thus vitalize the research on rare-earth dichalcogenides in the context of magnetization switching and magnetization dynamics.

\begin{acknowledgements}
This work was supported by the Federal Ministry of Education and Research of Germany in the framework of the Palestinian-German Science Bridge (BMBF grant number 01DH16027). We also gratefully acknowledge financial support by the Deutsche Forschungsgemeinschaft (DFG, German Research Foundation) $-$ TRR 288 $-$ 422213477 (project B06),  
%TRR 173/2 $-$ 268565370 (projects A11 and A01), 
CRC 1238 - 277146847 (Project C01), and the Sino-German research project DISTOMAT (MO 1731/10-1).  We  also gratefully acknowledge the J\"ulich Supercomputing Centre and RWTH Aachen University for providing computational resources under projects jiff40 and jara0062. 
%{\bf \color{red} jiff40?} 

\end{acknowledgements}

\bibliography{bib_Eu_Gd}

\begin{thebibliography}{48}%
\makeatletter
\providecommand \@ifxundefined [1]{%
 \@ifx{#1\undefined}
}%
\providecommand \@ifnum [1]{%
 \ifnum #1\expandafter \@firstoftwo
 \else \expandafter \@secondoftwo
 \fi
}%
\providecommand \@ifx [1]{%
 \ifx #1\expandafter \@firstoftwo
 \else \expandafter \@secondoftwo
 \fi
}%
\providecommand \natexlab [1]{#1}%
\providecommand \enquote  [1]{``#1''}%
\providecommand \bibnamefont  [1]{#1}%
\providecommand \bibfnamefont [1]{#1}%
\providecommand \citenamefont [1]{#1}%
\providecommand \href@noop [0]{\@secondoftwo}%
\providecommand \href [0]{\begingroup \@sanitize@url \@href}%
\providecommand \@href[1]{\@@startlink{#1}\@@href}%
\providecommand \@@href[1]{\endgroup#1\@@endlink}%
\providecommand \@sanitize@url [0]{\catcode `\\12\catcode `\$12\catcode
  `\&12\catcode `\#12\catcode `\^12\catcode `\_12\catcode `\%12\relax}%
\providecommand \@@startlink[1]{}%
\providecommand \@@endlink[0]{}%
\providecommand \url  [0]{\begingroup\@sanitize@url \@url }%
\providecommand \@url [1]{\endgroup\@href {#1}{\urlprefix }}%
\providecommand \urlprefix  [0]{URL }%
\providecommand \Eprint [0]{\href }%
\providecommand \doibase [0]{https://doi.org/}%
\providecommand \selectlanguage [0]{\@gobble}%
\providecommand \bibinfo  [0]{\@secondoftwo}%
\providecommand \bibfield  [0]{\@secondoftwo}%
\providecommand \translation [1]{[#1]}%
\providecommand \BibitemOpen [0]{}%
\providecommand \bibitemStop [0]{}%
\providecommand \bibitemNoStop [0]{.\EOS\space}%
\providecommand \EOS [0]{\spacefactor3000\relax}%
\providecommand \BibitemShut  [1]{\csname bibitem#1\endcsname}%
\let\auto@bib@innerbib\@empty
%</preamble>
\bibitem [{\citenamefont {Huang}\ \emph {et~al.}(2017)\citenamefont {Huang},
  \citenamefont {Clark}, \citenamefont {Navarro-Moratalla}, \citenamefont
  {Klein}, \citenamefont {Cheng}, \citenamefont {Seyler}, \citenamefont
  {Zhong}, \citenamefont {Schmidgall}, \citenamefont {McGuire}, \citenamefont
  {Cobden}, \citenamefont {Yao}, \citenamefont {Xiao}, \citenamefont
  {Jarillo-Herrero},\ and\ \citenamefont {Xu}}]{Huang2017}%
  \BibitemOpen
  \bibfield  {author} {\bibinfo {author} {\bibfnamefont {B.}~\bibnamefont
  {Huang}}, \bibinfo {author} {\bibfnamefont {G.}~\bibnamefont {Clark}},
  \bibinfo {author} {\bibfnamefont {E.}~\bibnamefont {Navarro-Moratalla}},
  \bibinfo {author} {\bibfnamefont {D.~R.}\ \bibnamefont {Klein}}, \bibinfo
  {author} {\bibfnamefont {R.}~\bibnamefont {Cheng}}, \bibinfo {author}
  {\bibfnamefont {K.~L.}\ \bibnamefont {Seyler}}, \bibinfo {author}
  {\bibfnamefont {D.}~\bibnamefont {Zhong}}, \bibinfo {author} {\bibfnamefont
  {E.}~\bibnamefont {Schmidgall}}, \bibinfo {author} {\bibfnamefont {M.~A.}\
  \bibnamefont {McGuire}}, \bibinfo {author} {\bibfnamefont {D.~H.}\
  \bibnamefont {Cobden}}, \bibinfo {author} {\bibfnamefont {W.}~\bibnamefont
  {Yao}}, \bibinfo {author} {\bibfnamefont {D.}~\bibnamefont {Xiao}}, \bibinfo
  {author} {\bibfnamefont {P.}~\bibnamefont {Jarillo-Herrero}},\ and\ \bibinfo
  {author} {\bibfnamefont {X.}~\bibnamefont {Xu}},\ }\bibfield  {title}
  {\bibinfo {title} {{Layer-dependent ferromagnetism in a van der Waals crystal
  down to the monolayer limit}},\ }\href {https://doi.org/10.1038/nature22391}
  {\bibfield  {journal} {\bibinfo  {journal} {Nature}\ }\textbf {\bibinfo
  {volume} {546}},\ \bibinfo {pages} {270} (\bibinfo {year}
  {2017})}\BibitemShut {NoStop}%
\bibitem [{\citenamefont {Wang}\ \emph
  {et~al.}(2020{\natexlab{a}})\citenamefont {Wang}, \citenamefont {Liu},
  \citenamefont {Wu}, \citenamefont {Hou}, \citenamefont {Jiang}, \citenamefont
  {Li}, \citenamefont {Pandey}, \citenamefont {Chen}, \citenamefont {Yang},
  \citenamefont {Wang}, \citenamefont {Wei}, \citenamefont {Lei}, \citenamefont
  {Kang}, \citenamefont {Wen}, \citenamefont {Nie}, \citenamefont {Zhao},\ and\
  \citenamefont {Wang}}]{Wang2020}%
  \BibitemOpen
  \bibfield  {author} {\bibinfo {author} {\bibfnamefont {H.}~\bibnamefont
  {Wang}}, \bibinfo {author} {\bibfnamefont {Y.}~\bibnamefont {Liu}}, \bibinfo
  {author} {\bibfnamefont {P.}~\bibnamefont {Wu}}, \bibinfo {author}
  {\bibfnamefont {W.}~\bibnamefont {Hou}}, \bibinfo {author} {\bibfnamefont
  {Y.}~\bibnamefont {Jiang}}, \bibinfo {author} {\bibfnamefont
  {X.}~\bibnamefont {Li}}, \bibinfo {author} {\bibfnamefont {C.}~\bibnamefont
  {Pandey}}, \bibinfo {author} {\bibfnamefont {D.}~\bibnamefont {Chen}},
  \bibinfo {author} {\bibfnamefont {Q.}~\bibnamefont {Yang}}, \bibinfo {author}
  {\bibfnamefont {H.}~\bibnamefont {Wang}}, \bibinfo {author} {\bibfnamefont
  {D.}~\bibnamefont {Wei}}, \bibinfo {author} {\bibfnamefont {N.}~\bibnamefont
  {Lei}}, \bibinfo {author} {\bibfnamefont {W.}~\bibnamefont {Kang}}, \bibinfo
  {author} {\bibfnamefont {L.}~\bibnamefont {Wen}}, \bibinfo {author}
  {\bibfnamefont {T.}~\bibnamefont {Nie}}, \bibinfo {author} {\bibfnamefont
  {W.}~\bibnamefont {Zhao}},\ and\ \bibinfo {author} {\bibfnamefont {K.~L.}\
  \bibnamefont {Wang}},\ }\bibfield  {title} {\bibinfo {title} {{Above
  Room-Temperature Ferromagnetism in Wafer-Scale Two-Dimensional van der Waals
  Fe3GeTe2 Tailored by a Topological Insulator}},\ }\href
  {https://doi.org/10.1021/acsnano.0c03152} {\bibfield  {journal} {\bibinfo
  {journal} {ACS Nano}\ }\textbf {\bibinfo {volume} {14}},\ \bibinfo {pages}
  {10045} (\bibinfo {year} {2020}{\natexlab{a}})},\ \bibinfo {note} {pMID:
  32686930},\ \Eprint
  {https://arxiv.org/abs/https://doi.org/10.1021/acsnano.0c03152}
  {https://doi.org/10.1021/acsnano.0c03152} \BibitemShut {NoStop}%
\bibitem [{\citenamefont {Olsen}(2021)}]{olsen2021}%
  \BibitemOpen
  \bibfield  {author} {\bibinfo {author} {\bibfnamefont {T.}~\bibnamefont
  {Olsen}},\ }\bibfield  {title} {\bibinfo {title} {{Magnetic anisotropy and
  exchange interactions of two-dimensional FePS3, NiPS3 and MnPS3 from first
  principles calculations}},\ }\href@noop {} {\bibfield  {journal} {\bibinfo
  {journal} {Journal of Physics D: Applied Physics}\ }\textbf {\bibinfo
  {volume} {54}},\ \bibinfo {pages} {314001} (\bibinfo {year}
  {2021})}\BibitemShut {NoStop}%
\bibitem [{\citenamefont {Wang}\ \emph
  {et~al.}(2020{\natexlab{b}})\citenamefont {Wang}, \citenamefont {Zhang},
  \citenamefont {Zhang}, \citenamefont {Yuan}, \citenamefont {Guo},
  \citenamefont {Dong},\ and\ \citenamefont {Wang}}]{wangGd2020}%
  \BibitemOpen
  \bibfield  {author} {\bibinfo {author} {\bibfnamefont {B.}~\bibnamefont
  {Wang}}, \bibinfo {author} {\bibfnamefont {X.}~\bibnamefont {Zhang}},
  \bibinfo {author} {\bibfnamefont {Y.}~\bibnamefont {Zhang}}, \bibinfo
  {author} {\bibfnamefont {S.}~\bibnamefont {Yuan}}, \bibinfo {author}
  {\bibfnamefont {Y.}~\bibnamefont {Guo}}, \bibinfo {author} {\bibfnamefont
  {S.}~\bibnamefont {Dong}},\ and\ \bibinfo {author} {\bibfnamefont
  {J.}~\bibnamefont {Wang}},\ }\bibfield  {title} {\bibinfo {title}
  {{Prediction of a two-dimensional high-TC f-electron ferromagnetic
  semiconductor}},\ }\href@noop {} {\bibfield  {journal} {\bibinfo  {journal}
  {Materials Horizons}\ }\textbf {\bibinfo {volume} {7}},\ \bibinfo {pages}
  {1623} (\bibinfo {year} {2020}{\natexlab{b}})}\BibitemShut {NoStop}%
\bibitem [{\citenamefont {Liu}\ \emph {et~al.}(2021)\citenamefont {Liu},
  \citenamefont {Tong}, \citenamefont {Deng}, \citenamefont {Yang},
  \citenamefont {Xie}, \citenamefont {Qin}, \citenamefont {Tian},\ and\
  \citenamefont {Zhang}}]{liu2021}%
  \BibitemOpen
  \bibfield  {author} {\bibinfo {author} {\bibfnamefont {W.}~\bibnamefont
  {Liu}}, \bibinfo {author} {\bibfnamefont {J.}~\bibnamefont {Tong}}, \bibinfo
  {author} {\bibfnamefont {L.}~\bibnamefont {Deng}}, \bibinfo {author}
  {\bibfnamefont {B.}~\bibnamefont {Yang}}, \bibinfo {author} {\bibfnamefont
  {G.}~\bibnamefont {Xie}}, \bibinfo {author} {\bibfnamefont {G.}~\bibnamefont
  {Qin}}, \bibinfo {author} {\bibfnamefont {F.}~\bibnamefont {Tian}},\ and\
  \bibinfo {author} {\bibfnamefont {X.}~\bibnamefont {Zhang}},\ }\bibfield
  {title} {\bibinfo {title} {{Two-dimensional ferromagnetic semiconductors of
  rare-earth monolayer GdX2 (X= Cl, Br, I) with large perpendicular magnetic
  anisotropy and high Curie temperature}},\ }\href@noop {} {\bibfield
  {journal} {\bibinfo  {journal} {Materials Today Physics}\ }\textbf {\bibinfo
  {volume} {21}},\ \bibinfo {pages} {100514} (\bibinfo {year}
  {2021})}\BibitemShut {NoStop}%
\bibitem [{\citenamefont {Sheng}\ \emph {et~al.}(2022)\citenamefont {Sheng},
  \citenamefont {Yuan},\ and\ \citenamefont {Wang}}]{sheng2022}%
  \BibitemOpen
  \bibfield  {author} {\bibinfo {author} {\bibfnamefont {K.}~\bibnamefont
  {Sheng}}, \bibinfo {author} {\bibfnamefont {H.-K.}\ \bibnamefont {Yuan}},\
  and\ \bibinfo {author} {\bibfnamefont {Z.-Y.}\ \bibnamefont {Wang}},\
  }\bibfield  {title} {\bibinfo {title} {{Monolayer gadolinium halides, GdX 2
  (X= F, Cl, Br): intrinsic ferrovalley materials with spontaneous spin and
  valley polarizations}},\ }\href@noop {} {\bibfield  {journal} {\bibinfo
  {journal} {Physical Chemistry Chemical Physics}\ }\textbf {\bibinfo {volume}
  {24}},\ \bibinfo {pages} {3865} (\bibinfo {year} {2022})}\BibitemShut
  {NoStop}%
\bibitem [{\citenamefont {You}\ \emph {et~al.}(2021)\citenamefont {You},
  \citenamefont {Zhang}, \citenamefont {Chen}, \citenamefont {Ding},
  \citenamefont {An}, \citenamefont {Miao},\ and\ \citenamefont
  {Dong}}]{you2021}%
  \BibitemOpen
  \bibfield  {author} {\bibinfo {author} {\bibfnamefont {H.}~\bibnamefont
  {You}}, \bibinfo {author} {\bibfnamefont {Y.}~\bibnamefont {Zhang}}, \bibinfo
  {author} {\bibfnamefont {J.}~\bibnamefont {Chen}}, \bibinfo {author}
  {\bibfnamefont {N.}~\bibnamefont {Ding}}, \bibinfo {author} {\bibfnamefont
  {M.}~\bibnamefont {An}}, \bibinfo {author} {\bibfnamefont {L.}~\bibnamefont
  {Miao}},\ and\ \bibinfo {author} {\bibfnamefont {S.}~\bibnamefont {Dong}},\
  }\bibfield  {title} {\bibinfo {title} {{:q! driven ferroelasticity in the
  two-dimensional d- f hybrid magnets}},\ }\href@noop {} {\bibfield  {journal}
  {\bibinfo  {journal} {Physical Review B}\ }\textbf {\bibinfo {volume}
  {103}},\ \bibinfo {pages} {L161408} (\bibinfo {year} {2021})}\BibitemShut
  {NoStop}%
\bibitem [{\citenamefont {Kato}\ \emph {et~al.}(2004)\citenamefont {Kato},
  \citenamefont {Myers}, \citenamefont {Gossard},\ and\ \citenamefont
  {Awschalom}}]{kato2004}%
  \BibitemOpen
  \bibfield  {author} {\bibinfo {author} {\bibfnamefont {Y.~K.}\ \bibnamefont
  {Kato}}, \bibinfo {author} {\bibfnamefont {R.~C.}\ \bibnamefont {Myers}},
  \bibinfo {author} {\bibfnamefont {A.~C.}\ \bibnamefont {Gossard}},\ and\
  \bibinfo {author} {\bibfnamefont {D.~D.}\ \bibnamefont {Awschalom}},\
  }\bibfield  {title} {\bibinfo {title} {{Observation of the spin Hall effect
  in semiconductors}},\ }\href@noop {} {\bibfield  {journal} {\bibinfo
  {journal} {science}\ }\textbf {\bibinfo {volume} {306}},\ \bibinfo {pages}
  {1910} (\bibinfo {year} {2004})}\BibitemShut {NoStop}%
\bibitem [{\citenamefont {S{\l}awi{\'n}ska}\ \emph {et~al.}(2019)\citenamefont
  {S{\l}awi{\'n}ska}, \citenamefont {Cerasoli}, \citenamefont {Wang},
  \citenamefont {Postorino}, \citenamefont {Supka}, \citenamefont {Curtarolo},
  \citenamefont {Fornari},\ and\ \citenamefont {Nardelli}}]{slawinska2019}%
  \BibitemOpen
  \bibfield  {author} {\bibinfo {author} {\bibfnamefont {J.}~\bibnamefont
  {S{\l}awi{\'n}ska}}, \bibinfo {author} {\bibfnamefont {F.~T.}\ \bibnamefont
  {Cerasoli}}, \bibinfo {author} {\bibfnamefont {H.}~\bibnamefont {Wang}},
  \bibinfo {author} {\bibfnamefont {S.}~\bibnamefont {Postorino}}, \bibinfo
  {author} {\bibfnamefont {A.}~\bibnamefont {Supka}}, \bibinfo {author}
  {\bibfnamefont {S.}~\bibnamefont {Curtarolo}}, \bibinfo {author}
  {\bibfnamefont {M.}~\bibnamefont {Fornari}},\ and\ \bibinfo {author}
  {\bibfnamefont {M.~B.}\ \bibnamefont {Nardelli}},\ }\bibfield  {title}
  {\bibinfo {title} {{Giant spin Hall effect in two-dimensional
  monochalcogenides}},\ }\href@noop {} {\bibfield  {journal} {\bibinfo
  {journal} {2D Materials}\ }\textbf {\bibinfo {volume} {6}},\ \bibinfo {pages}
  {025012} (\bibinfo {year} {2019})}\BibitemShut {NoStop}%
\bibitem [{\citenamefont {Safeer}\ \emph {et~al.}(2019)\citenamefont {Safeer},
  \citenamefont {Ingla-Ayn{\'e}s}, \citenamefont {Herling}, \citenamefont
  {Garcia}, \citenamefont {Vila}, \citenamefont {Ontoso}, \citenamefont
  {Calvo}, \citenamefont {Roche}, \citenamefont {Hueso},\ and\ \citenamefont
  {Casanova}}]{safeer2019}%
  \BibitemOpen
  \bibfield  {author} {\bibinfo {author} {\bibfnamefont {C.}~\bibnamefont
  {Safeer}}, \bibinfo {author} {\bibfnamefont {J.}~\bibnamefont
  {Ingla-Ayn{\'e}s}}, \bibinfo {author} {\bibfnamefont {F.}~\bibnamefont
  {Herling}}, \bibinfo {author} {\bibfnamefont {J.~H.}\ \bibnamefont {Garcia}},
  \bibinfo {author} {\bibfnamefont {M.}~\bibnamefont {Vila}}, \bibinfo {author}
  {\bibfnamefont {N.}~\bibnamefont {Ontoso}}, \bibinfo {author} {\bibfnamefont
  {M.~R.}\ \bibnamefont {Calvo}}, \bibinfo {author} {\bibfnamefont
  {S.}~\bibnamefont {Roche}}, \bibinfo {author} {\bibfnamefont {L.~E.}\
  \bibnamefont {Hueso}},\ and\ \bibinfo {author} {\bibfnamefont
  {F.}~\bibnamefont {Casanova}},\ }\bibfield  {title} {\bibinfo {title}
  {{Room-temperature spin Hall effect in graphene/MoS2 van der Waals
  heterostructures}},\ }\href@noop {} {\bibfield  {journal} {\bibinfo
  {journal} {Nano letters}\ }\textbf {\bibinfo {volume} {19}},\ \bibinfo
  {pages} {1074} (\bibinfo {year} {2019})}\BibitemShut {NoStop}%
\bibitem [{\citenamefont {Guo}\ \emph {et~al.}(2005)\citenamefont {Guo},
  \citenamefont {Yao},\ and\ \citenamefont {Niu}}]{guo2005}%
  \BibitemOpen
  \bibfield  {author} {\bibinfo {author} {\bibfnamefont {G.}~\bibnamefont
  {Guo}}, \bibinfo {author} {\bibfnamefont {Y.}~\bibnamefont {Yao}},\ and\
  \bibinfo {author} {\bibfnamefont {Q.}~\bibnamefont {Niu}},\ }\bibfield
  {title} {\bibinfo {title} {{Ab initio calculation of the intrinsic spin Hall
  effect in semiconductors}},\ }\href@noop {} {\bibfield  {journal} {\bibinfo
  {journal} {Physical review letters}\ }\textbf {\bibinfo {volume} {94}},\
  \bibinfo {pages} {226601} (\bibinfo {year} {2005})}\BibitemShut {NoStop}%
\bibitem [{\citenamefont {Go}\ \emph {et~al.}(2018)\citenamefont {Go},
  \citenamefont {Jo}, \citenamefont {Kim},\ and\ \citenamefont {Lee}}]{go2018}%
  \BibitemOpen
  \bibfield  {author} {\bibinfo {author} {\bibfnamefont {D.}~\bibnamefont
  {Go}}, \bibinfo {author} {\bibfnamefont {D.}~\bibnamefont {Jo}}, \bibinfo
  {author} {\bibfnamefont {C.}~\bibnamefont {Kim}},\ and\ \bibinfo {author}
  {\bibfnamefont {H.-W.}\ \bibnamefont {Lee}},\ }\bibfield  {title} {\bibinfo
  {title} {{Intrinsic spin and orbital Hall effects from orbital texture}},\
  }\href@noop {} {\bibfield  {journal} {\bibinfo  {journal} {Physical Review
  Letters}\ }\textbf {\bibinfo {volume} {121}},\ \bibinfo {pages} {086602}
  (\bibinfo {year} {2018})}\BibitemShut {NoStop}%
\bibitem [{\citenamefont {Go}\ \emph {et~al.}(2021)\citenamefont {Go},
  \citenamefont {Jo}, \citenamefont {Lee}, \citenamefont {Kl{\"a}ui},\ and\
  \citenamefont {Mokrousov}}]{go2021}%
  \BibitemOpen
  \bibfield  {author} {\bibinfo {author} {\bibfnamefont {D.}~\bibnamefont
  {Go}}, \bibinfo {author} {\bibfnamefont {D.}~\bibnamefont {Jo}}, \bibinfo
  {author} {\bibfnamefont {H.-W.}\ \bibnamefont {Lee}}, \bibinfo {author}
  {\bibfnamefont {M.}~\bibnamefont {Kl{\"a}ui}},\ and\ \bibinfo {author}
  {\bibfnamefont {Y.}~\bibnamefont {Mokrousov}},\ }\bibfield  {title} {\bibinfo
  {title} {{Orbitronics: Orbital currents in solids}},\ }\href@noop {}
  {\bibfield  {journal} {\bibinfo  {journal} {Europhysics Letters}\ }\textbf
  {\bibinfo {volume} {135}},\ \bibinfo {pages} {37001} (\bibinfo {year}
  {2021})}\BibitemShut {NoStop}%
\bibitem [{\citenamefont {Tanaka}\ \emph {et~al.}(2008)\citenamefont {Tanaka},
  \citenamefont {Kontani}, \citenamefont {Naito}, \citenamefont {Naito},
  \citenamefont {Hirashima}, \citenamefont {Yamada},\ and\ \citenamefont
  {Inoue}}]{tanaka2008}%
  \BibitemOpen
  \bibfield  {author} {\bibinfo {author} {\bibfnamefont {T.}~\bibnamefont
  {Tanaka}}, \bibinfo {author} {\bibfnamefont {H.}~\bibnamefont {Kontani}},
  \bibinfo {author} {\bibfnamefont {M.}~\bibnamefont {Naito}}, \bibinfo
  {author} {\bibfnamefont {T.}~\bibnamefont {Naito}}, \bibinfo {author}
  {\bibfnamefont {D.~S.}\ \bibnamefont {Hirashima}}, \bibinfo {author}
  {\bibfnamefont {K.}~\bibnamefont {Yamada}},\ and\ \bibinfo {author}
  {\bibfnamefont {J.}~\bibnamefont {Inoue}},\ }\bibfield  {title} {\bibinfo
  {title} {{Intrinsic spin Hall effect and orbital Hall effect in 4 d and 5 d
  transition metals}},\ }\href@noop {} {\bibfield  {journal} {\bibinfo
  {journal} {Physical Review B}\ }\textbf {\bibinfo {volume} {77}},\ \bibinfo
  {pages} {165117} (\bibinfo {year} {2008})}\BibitemShut {NoStop}%
\bibitem [{\citenamefont {Jo}\ \emph {et~al.}(2018)\citenamefont {Jo},
  \citenamefont {Go},\ and\ \citenamefont {Lee}}]{Jo2018}%
  \BibitemOpen
  \bibfield  {author} {\bibinfo {author} {\bibfnamefont {D.}~\bibnamefont
  {Jo}}, \bibinfo {author} {\bibfnamefont {D.}~\bibnamefont {Go}},\ and\
  \bibinfo {author} {\bibfnamefont {H.-W.}\ \bibnamefont {Lee}},\ }\bibfield
  {title} {\bibinfo {title} {{Gigantic intrinsic orbital Hall effects in weakly
  spin-orbit coupled metals}},\ }\href
  {https://doi.org/10.1103/PhysRevB.98.214405} {\bibfield  {journal} {\bibinfo
  {journal} {Phys. Rev. B}\ }\textbf {\bibinfo {volume} {98}},\ \bibinfo
  {pages} {214405} (\bibinfo {year} {2018})}\BibitemShut {NoStop}%
\bibitem [{\citenamefont {Canonico}\ \emph {et~al.}(2020)\citenamefont
  {Canonico}, \citenamefont {Cysne}, \citenamefont {Molina-Sanchez},
  \citenamefont {Muniz},\ and\ \citenamefont {Rappoport}}]{canonico2020}%
  \BibitemOpen
  \bibfield  {author} {\bibinfo {author} {\bibfnamefont {L.~M.}\ \bibnamefont
  {Canonico}}, \bibinfo {author} {\bibfnamefont {T.~P.}\ \bibnamefont {Cysne}},
  \bibinfo {author} {\bibfnamefont {A.}~\bibnamefont {Molina-Sanchez}},
  \bibinfo {author} {\bibfnamefont {R.}~\bibnamefont {Muniz}},\ and\ \bibinfo
  {author} {\bibfnamefont {T.~G.}\ \bibnamefont {Rappoport}},\ }\bibfield
  {title} {\bibinfo {title} {{Orbital Hall insulating phase in transition metal
  dichalcogenide monolayers}},\ }\href@noop {} {\bibfield  {journal} {\bibinfo
  {journal} {Physical Review B}\ }\textbf {\bibinfo {volume} {101}},\ \bibinfo
  {pages} {161409} (\bibinfo {year} {2020})}\BibitemShut {NoStop}%
\bibitem [{\citenamefont {Costa}\ \emph {et~al.}(2022)\citenamefont {Costa},
  \citenamefont {Focassio}, \citenamefont {Cysne}, \citenamefont {Canonico},
  \citenamefont {Schleder}, \citenamefont {Muniz}, \citenamefont {Fazzio},\
  and\ \citenamefont {Rappoport}}]{costa2022}%
  \BibitemOpen
  \bibfield  {author} {\bibinfo {author} {\bibfnamefont {M.}~\bibnamefont
  {Costa}}, \bibinfo {author} {\bibfnamefont {B.}~\bibnamefont {Focassio}},
  \bibinfo {author} {\bibfnamefont {T.~P.}\ \bibnamefont {Cysne}}, \bibinfo
  {author} {\bibfnamefont {L.~M.}\ \bibnamefont {Canonico}}, \bibinfo {author}
  {\bibfnamefont {G.~R.}\ \bibnamefont {Schleder}}, \bibinfo {author}
  {\bibfnamefont {R.~B.}\ \bibnamefont {Muniz}}, \bibinfo {author}
  {\bibfnamefont {A.}~\bibnamefont {Fazzio}},\ and\ \bibinfo {author}
  {\bibfnamefont {T.~G.}\ \bibnamefont {Rappoport}},\ }\bibfield  {title}
  {\bibinfo {title} {{Connecting Higher-Order Topology with the Orbital Hall
  Effect in Monolayers of Transition Metal Dichalcogenides}},\ }\href@noop {}
  {\bibfield  {journal} {\bibinfo  {journal} {arXiv preprint arXiv:2205.00997}\
  } (\bibinfo {year} {2022})}\BibitemShut {NoStop}%
\bibitem [{\citenamefont {Cysne}\ \emph {et~al.}(2021)\citenamefont {Cysne},
  \citenamefont {Costa}, \citenamefont {Canonico}, \citenamefont {Nardelli},
  \citenamefont {Muniz},\ and\ \citenamefont {Rappoport}}]{cysne2021}%
  \BibitemOpen
  \bibfield  {author} {\bibinfo {author} {\bibfnamefont {T.~P.}\ \bibnamefont
  {Cysne}}, \bibinfo {author} {\bibfnamefont {M.}~\bibnamefont {Costa}},
  \bibinfo {author} {\bibfnamefont {L.~M.}\ \bibnamefont {Canonico}}, \bibinfo
  {author} {\bibfnamefont {M.~B.}\ \bibnamefont {Nardelli}}, \bibinfo {author}
  {\bibfnamefont {R.}~\bibnamefont {Muniz}},\ and\ \bibinfo {author}
  {\bibfnamefont {T.~G.}\ \bibnamefont {Rappoport}},\ }\bibfield  {title}
  {\bibinfo {title} {{Disentangling orbital and valley Hall effects in bilayers
  of transition metal dichalcogenides}},\ }\href@noop {} {\bibfield  {journal}
  {\bibinfo  {journal} {Physical review letters}\ }\textbf {\bibinfo {volume}
  {126}},\ \bibinfo {pages} {056601} (\bibinfo {year} {2021})}\BibitemShut
  {NoStop}%
\bibitem [{\citenamefont {Bhowal}\ and\ \citenamefont
  {Satpathy}(2020)}]{Bhowal2020}%
  \BibitemOpen
  \bibfield  {author} {\bibinfo {author} {\bibfnamefont {S.}~\bibnamefont
  {Bhowal}}\ and\ \bibinfo {author} {\bibfnamefont {S.}~\bibnamefont
  {Satpathy}},\ }\bibfield  {title} {\bibinfo {title} {{Intrinsic orbital
  moment and prediction of a large orbital Hall effect in two-dimensional
  transition metal dichalcogenides}},\ }\href
  {https://doi.org/10.1103/PhysRevB.101.121112} {\bibfield  {journal} {\bibinfo
   {journal} {Phys. Rev. B}\ }\textbf {\bibinfo {volume} {101}},\ \bibinfo
  {pages} {121112} (\bibinfo {year} {2020})}\BibitemShut {NoStop}%
\bibitem [{\citenamefont {Go}\ \emph {et~al.}(2020)\citenamefont {Go},
  \citenamefont {Freimuth}, \citenamefont {Hanke}, \citenamefont {Xue},
  \citenamefont {Gomonay}, \citenamefont {Lee}, \citenamefont {Bl\"ugel},
  \citenamefont {Haney}, \citenamefont {Lee},\ and\ \citenamefont
  {Mokrousov}}]{Go2020}%
  \BibitemOpen
  \bibfield  {author} {\bibinfo {author} {\bibfnamefont {D.}~\bibnamefont
  {Go}}, \bibinfo {author} {\bibfnamefont {F.}~\bibnamefont {Freimuth}},
  \bibinfo {author} {\bibfnamefont {J.-P.}\ \bibnamefont {Hanke}}, \bibinfo
  {author} {\bibfnamefont {F.}~\bibnamefont {Xue}}, \bibinfo {author}
  {\bibfnamefont {O.}~\bibnamefont {Gomonay}}, \bibinfo {author} {\bibfnamefont
  {K.-J.}\ \bibnamefont {Lee}}, \bibinfo {author} {\bibfnamefont
  {S.}~\bibnamefont {Bl\"ugel}}, \bibinfo {author} {\bibfnamefont {P.~M.}\
  \bibnamefont {Haney}}, \bibinfo {author} {\bibfnamefont {H.-W.}\ \bibnamefont
  {Lee}},\ and\ \bibinfo {author} {\bibfnamefont {Y.}~\bibnamefont
  {Mokrousov}},\ }\bibfield  {title} {\bibinfo {title} {{Theory of
  current-induced angular momentum transfer dynamics in spin-orbit coupled
  systems}},\ }\href {https://doi.org/10.1103/PhysRevResearch.2.033401}
  {\bibfield  {journal} {\bibinfo  {journal} {Phys. Rev. Res.}\ }\textbf
  {\bibinfo {volume} {2}},\ \bibinfo {pages} {033401} (\bibinfo {year}
  {2020})}\BibitemShut {NoStop}%
\bibitem [{\citenamefont {Ramaswamy}\ \emph {et~al.}(2018)\citenamefont
  {Ramaswamy}, \citenamefont {Lee}, \citenamefont {Cai},\ and\ \citenamefont
  {Yang}}]{ramaswamy2018}%
  \BibitemOpen
  \bibfield  {author} {\bibinfo {author} {\bibfnamefont {R.}~\bibnamefont
  {Ramaswamy}}, \bibinfo {author} {\bibfnamefont {J.~M.}\ \bibnamefont {Lee}},
  \bibinfo {author} {\bibfnamefont {K.}~\bibnamefont {Cai}},\ and\ \bibinfo
  {author} {\bibfnamefont {H.}~\bibnamefont {Yang}},\ }\bibfield  {title}
  {\bibinfo {title} {{Recent advances in spin-orbit torques: Moving towards
  device applications}},\ }\href@noop {} {\bibfield  {journal} {\bibinfo
  {journal} {Applied Physics Reviews}\ }\textbf {\bibinfo {volume} {5}},\
  \bibinfo {pages} {031107} (\bibinfo {year} {2018})}\BibitemShut {NoStop}%
\bibitem [{\citenamefont {Go}\ and\ \citenamefont {Lee}(2020)}]{go2020orbital}%
  \BibitemOpen
  \bibfield  {author} {\bibinfo {author} {\bibfnamefont {D.}~\bibnamefont
  {Go}}\ and\ \bibinfo {author} {\bibfnamefont {H.-W.}\ \bibnamefont {Lee}},\
  }\bibfield  {title} {\bibinfo {title} {{Orbital torque: Torque generation by
  orbital current injection}},\ }\href@noop {} {\bibfield  {journal} {\bibinfo
  {journal} {Physical review research}\ }\textbf {\bibinfo {volume} {2}},\
  \bibinfo {pages} {013177} (\bibinfo {year} {2020})}\BibitemShut {NoStop}%
\bibitem [{\citenamefont {Smaili}\ \emph {et~al.}(2021)\citenamefont {Smaili},
  \citenamefont {Laref}, \citenamefont {Garcia}, \citenamefont
  {Schwingenschl{\"o}gl}, \citenamefont {Roche},\ and\ \citenamefont
  {Manchon}}]{smaili2021}%
  \BibitemOpen
  \bibfield  {author} {\bibinfo {author} {\bibfnamefont {I.}~\bibnamefont
  {Smaili}}, \bibinfo {author} {\bibfnamefont {S.}~\bibnamefont {Laref}},
  \bibinfo {author} {\bibfnamefont {J.~H.}\ \bibnamefont {Garcia}}, \bibinfo
  {author} {\bibfnamefont {U.}~\bibnamefont {Schwingenschl{\"o}gl}}, \bibinfo
  {author} {\bibfnamefont {S.}~\bibnamefont {Roche}},\ and\ \bibinfo {author}
  {\bibfnamefont {A.}~\bibnamefont {Manchon}},\ }\bibfield  {title} {\bibinfo
  {title} {{Janus monolayers of magnetic transition metal dichalcogenides as an
  all-in-one platform for spin-orbit torque}},\ }\href@noop {} {\bibfield
  {journal} {\bibinfo  {journal} {Physical Review B}\ }\textbf {\bibinfo
  {volume} {104}},\ \bibinfo {pages} {104415} (\bibinfo {year}
  {2021})}\BibitemShut {NoStop}%
\bibitem [{\citenamefont {Ding}\ \emph {et~al.}(2020)\citenamefont {Ding},
  \citenamefont {Ross}, \citenamefont {Go}, \citenamefont {Baldrati},
  \citenamefont {Ren}, \citenamefont {Freimuth}, \citenamefont {Becker},
  \citenamefont {Kammerbauer}, \citenamefont {Yang}, \citenamefont {Jakob}
  \emph {et~al.}}]{ding2020harnessing}%
  \BibitemOpen
  \bibfield  {author} {\bibinfo {author} {\bibfnamefont {S.}~\bibnamefont
  {Ding}}, \bibinfo {author} {\bibfnamefont {A.}~\bibnamefont {Ross}}, \bibinfo
  {author} {\bibfnamefont {D.}~\bibnamefont {Go}}, \bibinfo {author}
  {\bibfnamefont {L.}~\bibnamefont {Baldrati}}, \bibinfo {author}
  {\bibfnamefont {Z.}~\bibnamefont {Ren}}, \bibinfo {author} {\bibfnamefont
  {F.}~\bibnamefont {Freimuth}}, \bibinfo {author} {\bibfnamefont
  {S.}~\bibnamefont {Becker}}, \bibinfo {author} {\bibfnamefont
  {F.}~\bibnamefont {Kammerbauer}}, \bibinfo {author} {\bibfnamefont
  {J.}~\bibnamefont {Yang}}, \bibinfo {author} {\bibfnamefont {G.}~\bibnamefont
  {Jakob}}, \emph {et~al.},\ }\bibfield  {title} {\bibinfo {title} {{Harnessing
  orbital-to-spin conversion of interfacial orbital currents for efficient
  spin-orbit torques}},\ }\href@noop {} {\bibfield  {journal} {\bibinfo
  {journal} {Physical review letters}\ }\textbf {\bibinfo {volume} {125}},\
  \bibinfo {pages} {177201} (\bibinfo {year} {2020})}\BibitemShut {NoStop}%
\bibitem [{\citenamefont {Ding}\ \emph {et~al.}(2022)\citenamefont {Ding},
  \citenamefont {Liang}, \citenamefont {Go}, \citenamefont {Yun}, \citenamefont
  {Xue}, \citenamefont {Liu}, \citenamefont {Becker}, \citenamefont {Yang},
  \citenamefont {Du}, \citenamefont {Wang} \emph
  {et~al.}}]{ding2022observation}%
  \BibitemOpen
  \bibfield  {author} {\bibinfo {author} {\bibfnamefont {S.}~\bibnamefont
  {Ding}}, \bibinfo {author} {\bibfnamefont {Z.}~\bibnamefont {Liang}},
  \bibinfo {author} {\bibfnamefont {D.}~\bibnamefont {Go}}, \bibinfo {author}
  {\bibfnamefont {C.}~\bibnamefont {Yun}}, \bibinfo {author} {\bibfnamefont
  {M.}~\bibnamefont {Xue}}, \bibinfo {author} {\bibfnamefont {Z.}~\bibnamefont
  {Liu}}, \bibinfo {author} {\bibfnamefont {S.}~\bibnamefont {Becker}},
  \bibinfo {author} {\bibfnamefont {W.}~\bibnamefont {Yang}}, \bibinfo {author}
  {\bibfnamefont {H.}~\bibnamefont {Du}}, \bibinfo {author} {\bibfnamefont
  {C.}~\bibnamefont {Wang}}, \emph {et~al.},\ }\bibfield  {title} {\bibinfo
  {title} {{Observation of the orbital Rashba-Edelstein magnetoresistance}},\
  }\href@noop {} {\bibfield  {journal} {\bibinfo  {journal} {Physical review
  letters}\ }\textbf {\bibinfo {volume} {128}},\ \bibinfo {pages} {067201}
  (\bibinfo {year} {2022})}\BibitemShut {NoStop}%
\bibitem [{\citenamefont {Saunderson}\ \emph {et~al.}(2022)\citenamefont
  {Saunderson}, \citenamefont {Go}, \citenamefont {Bl\"ugel}, \citenamefont
  {Kl\"aui},\ and\ \citenamefont {Mokrousov}}]{Saunderson2022}%
  \BibitemOpen
  \bibfield  {author} {\bibinfo {author} {\bibfnamefont {T.~G.}\ \bibnamefont
  {Saunderson}}, \bibinfo {author} {\bibfnamefont {D.}~\bibnamefont {Go}},
  \bibinfo {author} {\bibfnamefont {S.}~\bibnamefont {Bl\"ugel}}, \bibinfo
  {author} {\bibfnamefont {M.}~\bibnamefont {Kl\"aui}},\ and\ \bibinfo {author}
  {\bibfnamefont {Y.}~\bibnamefont {Mokrousov}},\ }\bibfield  {title} {\bibinfo
  {title} {{Hidden interplay of current-induced spin and orbital torques in
  bulk ${\mathrm{Fe}}_{3}{\mathrm{GeTe}}_{2}$}},\ }\href
  {https://doi.org/10.1103/PhysRevResearch.4.L042022} {\bibfield  {journal}
  {\bibinfo  {journal} {Phys. Rev. Res.}\ }\textbf {\bibinfo {volume} {4}},\
  \bibinfo {pages} {L042022} (\bibinfo {year} {2022})}\BibitemShut {NoStop}%
\bibitem [{\citenamefont {Zeer}\ \emph {et~al.}(2022)\citenamefont {Zeer},
  \citenamefont {Go}, \citenamefont {Carbone}, \citenamefont {Saunderson},
  \citenamefont {Redies}, \citenamefont {Kl\"aui}, \citenamefont {Ghabboun},
  \citenamefont {Wulfhekel}, \citenamefont {Bl\"ugel},\ and\ \citenamefont
  {Mokrousov}}]{PhysRevMaterials.6.074004}%
  \BibitemOpen
  \bibfield  {author} {\bibinfo {author} {\bibfnamefont {M.}~\bibnamefont
  {Zeer}}, \bibinfo {author} {\bibfnamefont {D.}~\bibnamefont {Go}}, \bibinfo
  {author} {\bibfnamefont {J.~P.}\ \bibnamefont {Carbone}}, \bibinfo {author}
  {\bibfnamefont {T.~G.}\ \bibnamefont {Saunderson}}, \bibinfo {author}
  {\bibfnamefont {M.}~\bibnamefont {Redies}}, \bibinfo {author} {\bibfnamefont
  {M.}~\bibnamefont {Kl\"aui}}, \bibinfo {author} {\bibfnamefont
  {J.}~\bibnamefont {Ghabboun}}, \bibinfo {author} {\bibfnamefont
  {W.}~\bibnamefont {Wulfhekel}}, \bibinfo {author} {\bibfnamefont
  {S.}~\bibnamefont {Bl\"ugel}},\ and\ \bibinfo {author} {\bibfnamefont
  {Y.}~\bibnamefont {Mokrousov}},\ }\bibfield  {title} {\bibinfo {title} {{Spin
  and orbital transport in rare-earth dichalcogenides: The case of
  ${\mathrm{EuS}}_{2}$}},\ }\href
  {https://doi.org/10.1103/PhysRevMaterials.6.074004} {\bibfield  {journal}
  {\bibinfo  {journal} {Phys. Rev. Mater.}\ }\textbf {\bibinfo {volume} {6}},\
  \bibinfo {pages} {074004} (\bibinfo {year} {2022})}\BibitemShut {NoStop}%
\bibitem [{\citenamefont {Wortmann}\ \emph {et~al.}(2023)\citenamefont
  {Wortmann}, \citenamefont {Kl{\"u}ppelberg}, \citenamefont {Bihlmayer},
  \citenamefont {Heide}, \citenamefont {Madsen}, \citenamefont {Friedrich},
  \citenamefont {Winkelmann}, \citenamefont {Schlipf}, \citenamefont {Hilgers},
  \citenamefont {Kurz} \emph {et~al.}}]{wortmann2023fleur}%
  \BibitemOpen
  \bibfield  {author} {\bibinfo {author} {\bibfnamefont {D.}~\bibnamefont
  {Wortmann}}, \bibinfo {author} {\bibfnamefont {D.~A.}\ \bibnamefont
  {Kl{\"u}ppelberg}}, \bibinfo {author} {\bibfnamefont {G.}~\bibnamefont
  {Bihlmayer}}, \bibinfo {author} {\bibfnamefont {M.}~\bibnamefont {Heide}},
  \bibinfo {author} {\bibfnamefont {G.~K.}\ \bibnamefont {Madsen}}, \bibinfo
  {author} {\bibfnamefont {C.}~\bibnamefont {Friedrich}}, \bibinfo {author}
  {\bibfnamefont {M.}~\bibnamefont {Winkelmann}}, \bibinfo {author}
  {\bibfnamefont {M.}~\bibnamefont {Schlipf}}, \bibinfo {author} {\bibfnamefont
  {R.}~\bibnamefont {Hilgers}}, \bibinfo {author} {\bibfnamefont
  {P.}~\bibnamefont {Kurz}}, \emph {et~al.},\ }\href@noop {} {\emph {\bibinfo
  {title} {Fleur}}},\ \bibinfo {type} {Tech. Rep.}\ (\bibinfo  {institution}
  {Quanten-Theorie der Materialien},\ \bibinfo {year} {2023})\BibitemShut
  {NoStop}%
\bibitem [{\citenamefont {Wimmer}\ \emph {et~al.}(1981)\citenamefont {Wimmer},
  \citenamefont {Krakauer}, \citenamefont {Weinert},\ and\ \citenamefont
  {Freeman}}]{Wimmer1981}%
  \BibitemOpen
  \bibfield  {author} {\bibinfo {author} {\bibfnamefont {E.}~\bibnamefont
  {Wimmer}}, \bibinfo {author} {\bibfnamefont {H.}~\bibnamefont {Krakauer}},
  \bibinfo {author} {\bibfnamefont {M.}~\bibnamefont {Weinert}},\ and\ \bibinfo
  {author} {\bibfnamefont {A.~J.}\ \bibnamefont {Freeman}},\ }\bibfield
  {title} {\bibinfo {title} {{Full-potential self-consistent
  linearized-augmented-plane-wave method for calculating the electronic
  structure of molecules and surfaces: $\mathrm{O}_{2}$ molecule}},\ }\href
  {https://doi.org/10.1103/PhysRevB.24.864} {\bibfield  {journal} {\bibinfo
  {journal} {Phys. Rev. B}\ }\textbf {\bibinfo {volume} {24}},\ \bibinfo
  {pages} {864} (\bibinfo {year} {1981})}\BibitemShut {NoStop}%
\bibitem [{\citenamefont {Perdew}\ \emph {et~al.}(1996)\citenamefont {Perdew},
  \citenamefont {Burke},\ and\ \citenamefont {Ernzerhof}}]{Perdew1996}%
  \BibitemOpen
  \bibfield  {author} {\bibinfo {author} {\bibfnamefont {J.~P.}\ \bibnamefont
  {Perdew}}, \bibinfo {author} {\bibfnamefont {K.}~\bibnamefont {Burke}},\ and\
  \bibinfo {author} {\bibfnamefont {M.}~\bibnamefont {Ernzerhof}},\ }\bibfield
  {title} {\bibinfo {title} {{Generalized Gradient Approximation Made
  Simple}},\ }\href {https://doi.org/10.1103/PhysRevLett.77.3865} {\bibfield
  {journal} {\bibinfo  {journal} {Phys. Rev. Lett.}\ }\textbf {\bibinfo
  {volume} {77}},\ \bibinfo {pages} {3865} (\bibinfo {year}
  {1996})}\BibitemShut {NoStop}%
\bibitem [{\citenamefont {Pizzi}\ \emph {et~al.}(2020)\citenamefont {Pizzi},
  \citenamefont {Vitale}, \citenamefont {Arita}, \citenamefont {Bl{\"u}gel},
  \citenamefont {Freimuth}, \citenamefont {G{\'e}ranton}, \citenamefont
  {Gibertini}, \citenamefont {Gresch}, \citenamefont {Johnson}, \citenamefont
  {Koretsune} \emph {et~al.}}]{pizzi2020wannier90}%
  \BibitemOpen
  \bibfield  {author} {\bibinfo {author} {\bibfnamefont {G.}~\bibnamefont
  {Pizzi}}, \bibinfo {author} {\bibfnamefont {V.}~\bibnamefont {Vitale}},
  \bibinfo {author} {\bibfnamefont {R.}~\bibnamefont {Arita}}, \bibinfo
  {author} {\bibfnamefont {S.}~\bibnamefont {Bl{\"u}gel}}, \bibinfo {author}
  {\bibfnamefont {F.}~\bibnamefont {Freimuth}}, \bibinfo {author}
  {\bibfnamefont {G.}~\bibnamefont {G{\'e}ranton}}, \bibinfo {author}
  {\bibfnamefont {M.}~\bibnamefont {Gibertini}}, \bibinfo {author}
  {\bibfnamefont {D.}~\bibnamefont {Gresch}}, \bibinfo {author} {\bibfnamefont
  {C.}~\bibnamefont {Johnson}}, \bibinfo {author} {\bibfnamefont
  {T.}~\bibnamefont {Koretsune}}, \emph {et~al.},\ }\bibfield  {title}
  {\bibinfo {title} {{Wannier90 as a community code: new features and
  applications}},\ }\href@noop {} {\bibfield  {journal} {\bibinfo  {journal}
  {Journal of Physics: Condensed Matter}\ }\textbf {\bibinfo {volume} {32}},\
  \bibinfo {pages} {165902} (\bibinfo {year} {2020})}\BibitemShut {NoStop}%
\bibitem [{\citenamefont {Shick}\ \emph {et~al.}(1999)\citenamefont {Shick},
  \citenamefont {Liechtenstein},\ and\ \citenamefont {Pickett}}]{Shick1999}%
  \BibitemOpen
  \bibfield  {author} {\bibinfo {author} {\bibfnamefont {A.~B.}\ \bibnamefont
  {Shick}}, \bibinfo {author} {\bibfnamefont {A.~I.}\ \bibnamefont
  {Liechtenstein}},\ and\ \bibinfo {author} {\bibfnamefont {W.~E.}\
  \bibnamefont {Pickett}},\ }\bibfield  {title} {\bibinfo {title}
  {{Implementation of the $\text{LDA+U}$ method using the full-potential
  linearized augmented plane-wave basis}},\ }\href
  {https://doi.org/10.1103/PhysRevB.60.10763} {\bibfield  {journal} {\bibinfo
  {journal} {Phys. Rev. B}\ }\textbf {\bibinfo {volume} {60}},\ \bibinfo
  {pages} {10763} (\bibinfo {year} {1999})}\BibitemShut {NoStop}%
\bibitem [{\citenamefont {Kurz}\ \emph {et~al.}(2002)\citenamefont {Kurz},
  \citenamefont {Bihlmayer},\ and\ \citenamefont {Bl{\"u}gel}}]{Kurz2002}%
  \BibitemOpen
  \bibfield  {author} {\bibinfo {author} {\bibfnamefont {P.}~\bibnamefont
  {Kurz}}, \bibinfo {author} {\bibfnamefont {G.}~\bibnamefont {Bihlmayer}},\
  and\ \bibinfo {author} {\bibfnamefont {S.}~\bibnamefont {Bl{\"u}gel}},\
  }\bibfield  {title} {\bibinfo {title} {{Magnetism and electronic structure of
  hcp Gd and the Gd(0001) surface}},\ }\href@noop {} {\bibfield  {journal}
  {\bibinfo  {journal} {Journal of Physics: Condensed Matter}\ }\textbf
  {\bibinfo {volume} {14}},\ \bibinfo {pages} {6353} (\bibinfo {year}
  {2002})}\BibitemShut {NoStop}%
\bibitem [{\citenamefont {Carbone}\ \emph {et~al.}(2022)\citenamefont
  {Carbone}, \citenamefont {Go}, \citenamefont {Mokrousov}, \citenamefont
  {Bihlmayer},\ and\ \citenamefont {Bl\"ugel}}]{carbone2022}%
  \BibitemOpen
  \bibfield  {author} {\bibinfo {author} {\bibfnamefont {J.~P.}\ \bibnamefont
  {Carbone}}, \bibinfo {author} {\bibfnamefont {D.}~\bibnamefont {Go}},
  \bibinfo {author} {\bibfnamefont {Y.}~\bibnamefont {Mokrousov}}, \bibinfo
  {author} {\bibfnamefont {G.}~\bibnamefont {Bihlmayer}},\ and\ \bibinfo
  {author} {\bibfnamefont {S.}~\bibnamefont {Bl\"ugel}},\ }\bibfield  {title}
  {\bibinfo {title} {{Engineering spin-orbit effects and Berry curvature by
  deposition of a monolayer of Eu on ${\mathrm{WSe}}_{2}$}},\ }\href
  {https://doi.org/10.1103/PhysRevB.106.064401} {\bibfield  {journal} {\bibinfo
   {journal} {Phys. Rev. B}\ }\textbf {\bibinfo {volume} {106}},\ \bibinfo
  {pages} {064401} (\bibinfo {year} {2022})}\BibitemShut {NoStop}%
\bibitem [{\citenamefont {Li}\ \emph {et~al.}(2012)\citenamefont {Li},
  \citenamefont {Wu}, \citenamefont {Li}, \citenamefont {Yang},\ and\
  \citenamefont {Hou}}]{li2012}%
  \BibitemOpen
  \bibfield  {author} {\bibinfo {author} {\bibfnamefont {X.}~\bibnamefont
  {Li}}, \bibinfo {author} {\bibfnamefont {X.}~\bibnamefont {Wu}}, \bibinfo
  {author} {\bibfnamefont {Z.}~\bibnamefont {Li}}, \bibinfo {author}
  {\bibfnamefont {J.}~\bibnamefont {Yang}},\ and\ \bibinfo {author}
  {\bibfnamefont {J.}~\bibnamefont {Hou}},\ }\bibfield  {title} {\bibinfo
  {title} {{Bipolar magnetic semiconductors: a new class of spintronics
  materials}},\ }\href@noop {} {\bibfield  {journal} {\bibinfo  {journal}
  {Nanoscale}\ }\textbf {\bibinfo {volume} {4}},\ \bibinfo {pages} {5680}
  (\bibinfo {year} {2012})}\BibitemShut {NoStop}%
\bibitem [{\citenamefont {Cheng}\ \emph {et~al.}(2018)\citenamefont {Cheng},
  \citenamefont {Zhou}, \citenamefont {Yang}, \citenamefont {Shen},
  \citenamefont {Linghu}, \citenamefont {Wu}, \citenamefont {Qian},\ and\
  \citenamefont {Feng}}]{cheng2018}%
  \BibitemOpen
  \bibfield  {author} {\bibinfo {author} {\bibfnamefont {H.}~\bibnamefont
  {Cheng}}, \bibinfo {author} {\bibfnamefont {J.}~\bibnamefont {Zhou}},
  \bibinfo {author} {\bibfnamefont {M.}~\bibnamefont {Yang}}, \bibinfo {author}
  {\bibfnamefont {L.}~\bibnamefont {Shen}}, \bibinfo {author} {\bibfnamefont
  {J.}~\bibnamefont {Linghu}}, \bibinfo {author} {\bibfnamefont
  {Q.}~\bibnamefont {Wu}}, \bibinfo {author} {\bibfnamefont {P.}~\bibnamefont
  {Qian}},\ and\ \bibinfo {author} {\bibfnamefont {Y.~P.}\ \bibnamefont
  {Feng}},\ }\bibfield  {title} {\bibinfo {title} {{Robust two-dimensional
  bipolar magnetic semiconductors by defect engineering}},\ }\href@noop {}
  {\bibfield  {journal} {\bibinfo  {journal} {Journal of Materials Chemistry
  C}\ }\textbf {\bibinfo {volume} {6}},\ \bibinfo {pages} {8435} (\bibinfo
  {year} {2018})}\BibitemShut {NoStop}%
\bibitem [{\citenamefont {Sancho}\ \emph {et~al.}(1985)\citenamefont {Sancho},
  \citenamefont {Sancho}, \citenamefont {Sancho},\ and\ \citenamefont
  {Rubio}}]{sancho1985highly}%
  \BibitemOpen
  \bibfield  {author} {\bibinfo {author} {\bibfnamefont {M.~L.}\ \bibnamefont
  {Sancho}}, \bibinfo {author} {\bibfnamefont {J.~L.}\ \bibnamefont {Sancho}},
  \bibinfo {author} {\bibfnamefont {J.~L.}\ \bibnamefont {Sancho}},\ and\
  \bibinfo {author} {\bibfnamefont {J.}~\bibnamefont {Rubio}},\ }\bibfield
  {title} {\bibinfo {title} {{Highly convergent schemes for the calculation of
  bulk and surface Green functions}},\ }\href@noop {} {\bibfield  {journal}
  {\bibinfo  {journal} {Journal of Physics F: Metal Physics}\ }\textbf
  {\bibinfo {volume} {15}},\ \bibinfo {pages} {851} (\bibinfo {year}
  {1985})}\BibitemShut {NoStop}%
\bibitem [{\citenamefont {Wu}\ \emph {et~al.}(2018)\citenamefont {Wu},
  \citenamefont {Zhang}, \citenamefont {Song}, \citenamefont {Troyer},\ and\
  \citenamefont {Soluyanov}}]{wu2018wanniertools}%
  \BibitemOpen
  \bibfield  {author} {\bibinfo {author} {\bibfnamefont {Q.}~\bibnamefont
  {Wu}}, \bibinfo {author} {\bibfnamefont {S.}~\bibnamefont {Zhang}}, \bibinfo
  {author} {\bibfnamefont {H.-F.}\ \bibnamefont {Song}}, \bibinfo {author}
  {\bibfnamefont {M.}~\bibnamefont {Troyer}},\ and\ \bibinfo {author}
  {\bibfnamefont {A.~A.}\ \bibnamefont {Soluyanov}},\ }\bibfield  {title}
  {\bibinfo {title} {{WannierTools: An open-source software package for novel
  topological materials}},\ }\href@noop {} {\bibfield  {journal} {\bibinfo
  {journal} {Computer Physics Communications}\ }\textbf {\bibinfo {volume}
  {224}},\ \bibinfo {pages} {405} (\bibinfo {year} {2018})}\BibitemShut
  {NoStop}%
\bibitem [{\citenamefont {Kontani}\ \emph {et~al.}(2009)\citenamefont
  {Kontani}, \citenamefont {Tanaka}, \citenamefont {Hirashima}, \citenamefont
  {Yamada},\ and\ \citenamefont {Inoue}}]{kontani2009}%
  \BibitemOpen
  \bibfield  {author} {\bibinfo {author} {\bibfnamefont {H.}~\bibnamefont
  {Kontani}}, \bibinfo {author} {\bibfnamefont {T.}~\bibnamefont {Tanaka}},
  \bibinfo {author} {\bibfnamefont {D.~S.}\ \bibnamefont {Hirashima}}, \bibinfo
  {author} {\bibfnamefont {K.}~\bibnamefont {Yamada}},\ and\ \bibinfo {author}
  {\bibfnamefont {J.}~\bibnamefont {Inoue}},\ }\bibfield  {title} {\bibinfo
  {title} {{Giant Orbital Hall Effect in Transition Metals: Origin of Large
  Spin and Anomalous Hall Effects}},\ }\href@noop {} {\bibfield  {journal}
  {\bibinfo  {journal} {Physical Review Letters}\ }\textbf {\bibinfo {volume}
  {102}},\ \bibinfo {pages} {016601} (\bibinfo {year} {2009})}\BibitemShut
  {NoStop}%
\bibitem [{\citenamefont {Ekimov}\ \emph {et~al.}(2023)\citenamefont {Ekimov},
  \citenamefont {Nikolaev}, \citenamefont {Ivanova}, \citenamefont {Sidorov},
  \citenamefont {Shiryaev}, \citenamefont {Usmanov}, \citenamefont {Vasiliev},
  \citenamefont {Artemov}, \citenamefont {Kondrin}, \citenamefont
  {Chernopitsskiy} \emph {et~al.}}]{ekimov2023}%
  \BibitemOpen
  \bibfield  {author} {\bibinfo {author} {\bibfnamefont {E.}~\bibnamefont
  {Ekimov}}, \bibinfo {author} {\bibfnamefont {S.}~\bibnamefont {Nikolaev}},
  \bibinfo {author} {\bibfnamefont {A.}~\bibnamefont {Ivanova}}, \bibinfo
  {author} {\bibfnamefont {V.}~\bibnamefont {Sidorov}}, \bibinfo {author}
  {\bibfnamefont {A.}~\bibnamefont {Shiryaev}}, \bibinfo {author}
  {\bibfnamefont {I.}~\bibnamefont {Usmanov}}, \bibinfo {author} {\bibfnamefont
  {A.}~\bibnamefont {Vasiliev}}, \bibinfo {author} {\bibfnamefont
  {V.}~\bibnamefont {Artemov}}, \bibinfo {author} {\bibfnamefont
  {M.}~\bibnamefont {Kondrin}}, \bibinfo {author} {\bibfnamefont
  {M.}~\bibnamefont {Chernopitsskiy}}, \emph {et~al.},\ }\bibfield  {title}
  {\bibinfo {title} {{Structural, optical and transport properties of layered
  europium disulfide synthesized under high pressure}},\ }\href@noop {}
  {\bibfield  {journal} {\bibinfo  {journal} {CrystEngComm}\ }\textbf {\bibinfo
  {volume} {25}},\ \bibinfo {pages} {2966} (\bibinfo {year}
  {2023})}\BibitemShut {NoStop}%
\bibitem [{\citenamefont {M{\l}y{\'n}czak}\ \emph {et~al.}(2022)\citenamefont
  {M{\l}y{\'n}czak}, \citenamefont {Aguilera}, \citenamefont
  {Gospodari{\v{c}}}, \citenamefont {Heider}, \citenamefont {Jugovac},
  \citenamefont {Zamborlini}, \citenamefont {Hanke}, \citenamefont {Friedrich},
  \citenamefont {Mokrousov}, \citenamefont {Tusche} \emph
  {et~al.}}]{mlynczak2022fe}%
  \BibitemOpen
  \bibfield  {author} {\bibinfo {author} {\bibfnamefont {E.}~\bibnamefont
  {M{\l}y{\'n}czak}}, \bibinfo {author} {\bibfnamefont {I.}~\bibnamefont
  {Aguilera}}, \bibinfo {author} {\bibfnamefont {P.}~\bibnamefont
  {Gospodari{\v{c}}}}, \bibinfo {author} {\bibfnamefont {T.}~\bibnamefont
  {Heider}}, \bibinfo {author} {\bibfnamefont {M.}~\bibnamefont {Jugovac}},
  \bibinfo {author} {\bibfnamefont {G.}~\bibnamefont {Zamborlini}}, \bibinfo
  {author} {\bibfnamefont {J.-P.}\ \bibnamefont {Hanke}}, \bibinfo {author}
  {\bibfnamefont {C.}~\bibnamefont {Friedrich}}, \bibinfo {author}
  {\bibfnamefont {Y.}~\bibnamefont {Mokrousov}}, \bibinfo {author}
  {\bibfnamefont {C.}~\bibnamefont {Tusche}}, \emph {et~al.},\ }\bibfield
  {title} {\bibinfo {title} {{Fe (001) angle-resolved photoemission and
  intrinsic anomalous Hall conductivity in Fe seen by different ab initio
  approaches: LDA and GGA versus GW}},\ }\href@noop {} {\bibfield  {journal}
  {\bibinfo  {journal} {Physical Review B}\ }\textbf {\bibinfo {volume}
  {105}},\ \bibinfo {pages} {115135} (\bibinfo {year} {2022})}\BibitemShut
  {NoStop}%
\bibitem [{\citenamefont {Chen}\ \emph
  {et~al.}(2022{\natexlab{a}})\citenamefont {Chen}, \citenamefont {Hanke},
  \citenamefont {Hoffmann}, \citenamefont {Bihlmayer}, \citenamefont
  {Mokrousov}, \citenamefont {Bl{\"u}gel}, \citenamefont {Schneider},\ and\
  \citenamefont {Tusche}}]{chen2022spanning}%
  \BibitemOpen
  \bibfield  {author} {\bibinfo {author} {\bibfnamefont {Y.-J.}\ \bibnamefont
  {Chen}}, \bibinfo {author} {\bibfnamefont {J.-P.}\ \bibnamefont {Hanke}},
  \bibinfo {author} {\bibfnamefont {M.}~\bibnamefont {Hoffmann}}, \bibinfo
  {author} {\bibfnamefont {G.}~\bibnamefont {Bihlmayer}}, \bibinfo {author}
  {\bibfnamefont {Y.}~\bibnamefont {Mokrousov}}, \bibinfo {author}
  {\bibfnamefont {S.}~\bibnamefont {Bl{\"u}gel}}, \bibinfo {author}
  {\bibfnamefont {C.~M.}\ \bibnamefont {Schneider}},\ and\ \bibinfo {author}
  {\bibfnamefont {C.}~\bibnamefont {Tusche}},\ }\bibfield  {title} {\bibinfo
  {title} {{Spanning Fermi arcs in a two-dimensional magnet}},\ }\href@noop {}
  {\bibfield  {journal} {\bibinfo  {journal} {Nature Communications}\ }\textbf
  {\bibinfo {volume} {13}},\ \bibinfo {pages} {5309} (\bibinfo {year}
  {2022}{\natexlab{a}})}\BibitemShut {NoStop}%
\bibitem [{\citenamefont {Niu}\ \emph {et~al.}(2019)\citenamefont {Niu},
  \citenamefont {Hanke}, \citenamefont {Buhl}, \citenamefont {Zhang},
  \citenamefont {Plucinski}, \citenamefont {Wortmann}, \citenamefont
  {Bl{\"u}gel}, \citenamefont {Bihlmayer},\ and\ \citenamefont
  {Mokrousov}}]{niu2019mixed}%
  \BibitemOpen
  \bibfield  {author} {\bibinfo {author} {\bibfnamefont {C.}~\bibnamefont
  {Niu}}, \bibinfo {author} {\bibfnamefont {J.-P.}\ \bibnamefont {Hanke}},
  \bibinfo {author} {\bibfnamefont {P.~M.}\ \bibnamefont {Buhl}}, \bibinfo
  {author} {\bibfnamefont {H.}~\bibnamefont {Zhang}}, \bibinfo {author}
  {\bibfnamefont {L.}~\bibnamefont {Plucinski}}, \bibinfo {author}
  {\bibfnamefont {D.}~\bibnamefont {Wortmann}}, \bibinfo {author}
  {\bibfnamefont {S.}~\bibnamefont {Bl{\"u}gel}}, \bibinfo {author}
  {\bibfnamefont {G.}~\bibnamefont {Bihlmayer}},\ and\ \bibinfo {author}
  {\bibfnamefont {Y.}~\bibnamefont {Mokrousov}},\ }\bibfield  {title} {\bibinfo
  {title} {{Mixed topological semimetals driven by orbital complexity in
  two-dimensional ferromagnets}},\ }\href@noop {} {\bibfield  {journal}
  {\bibinfo  {journal} {Nature communications}\ }\textbf {\bibinfo {volume}
  {10}},\ \bibinfo {pages} {3179} (\bibinfo {year} {2019})}\BibitemShut
  {NoStop}%
\bibitem [{\citenamefont {Hanke}\ \emph {et~al.}(2017)\citenamefont {Hanke},
  \citenamefont {Freimuth}, \citenamefont {Niu}, \citenamefont {Bl{\"u}gel},\
  and\ \citenamefont {Mokrousov}}]{hanke2017mixed}%
  \BibitemOpen
  \bibfield  {author} {\bibinfo {author} {\bibfnamefont {J.-P.}\ \bibnamefont
  {Hanke}}, \bibinfo {author} {\bibfnamefont {F.}~\bibnamefont {Freimuth}},
  \bibinfo {author} {\bibfnamefont {C.}~\bibnamefont {Niu}}, \bibinfo {author}
  {\bibfnamefont {S.}~\bibnamefont {Bl{\"u}gel}},\ and\ \bibinfo {author}
  {\bibfnamefont {Y.}~\bibnamefont {Mokrousov}},\ }\bibfield  {title} {\bibinfo
  {title} {{Mixed Weyl semimetals and low-dissipation magnetization control in
  insulators by spin--orbit torques}},\ }\href@noop {} {\bibfield  {journal}
  {\bibinfo  {journal} {Nature Communications}\ }\textbf {\bibinfo {volume}
  {8}},\ \bibinfo {pages} {1479} (\bibinfo {year} {2017})}\BibitemShut
  {NoStop}%
\bibitem [{\citenamefont {Liu}\ \emph {et~al.}(2022)\citenamefont {Liu},
  \citenamefont {Li}, \citenamefont {Zhang},\ and\ \citenamefont
  {Yan}}]{liu2022janus}%
  \BibitemOpen
  \bibfield  {author} {\bibinfo {author} {\bibfnamefont {W.}~\bibnamefont
  {Liu}}, \bibinfo {author} {\bibfnamefont {X.}~\bibnamefont {Li}}, \bibinfo
  {author} {\bibfnamefont {C.}~\bibnamefont {Zhang}},\ and\ \bibinfo {author}
  {\bibfnamefont {S.}~\bibnamefont {Yan}},\ }\bibfield  {title} {\bibinfo
  {title} {{Janus VXY monolayers with tunable large Berry curvature}},\
  }\href@noop {} {\bibfield  {journal} {\bibinfo  {journal} {Journal of
  Semiconductors}\ }\textbf {\bibinfo {volume} {43}},\ \bibinfo {pages}
  {042501} (\bibinfo {year} {2022})}\BibitemShut {NoStop}%
\bibitem [{\citenamefont {Lu}\ \emph {et~al.}(2017)\citenamefont {Lu},
  \citenamefont {Zhu}, \citenamefont {Xiao}, \citenamefont {Chuu},
  \citenamefont {Han}, \citenamefont {Chiu}, \citenamefont {Cheng},
  \citenamefont {Yang}, \citenamefont {Wei}, \citenamefont {Yang} \emph
  {et~al.}}]{lu2017janus}%
  \BibitemOpen
  \bibfield  {author} {\bibinfo {author} {\bibfnamefont {A.-Y.}\ \bibnamefont
  {Lu}}, \bibinfo {author} {\bibfnamefont {H.}~\bibnamefont {Zhu}}, \bibinfo
  {author} {\bibfnamefont {J.}~\bibnamefont {Xiao}}, \bibinfo {author}
  {\bibfnamefont {C.-P.}\ \bibnamefont {Chuu}}, \bibinfo {author}
  {\bibfnamefont {Y.}~\bibnamefont {Han}}, \bibinfo {author} {\bibfnamefont
  {M.-H.}\ \bibnamefont {Chiu}}, \bibinfo {author} {\bibfnamefont {C.-C.}\
  \bibnamefont {Cheng}}, \bibinfo {author} {\bibfnamefont {C.-W.}\ \bibnamefont
  {Yang}}, \bibinfo {author} {\bibfnamefont {K.-H.}\ \bibnamefont {Wei}},
  \bibinfo {author} {\bibfnamefont {Y.}~\bibnamefont {Yang}}, \emph {et~al.},\
  }\bibfield  {title} {\bibinfo {title} {{Janus monolayers of transition metal
  dichalcogenides}},\ }\href@noop {} {\bibfield  {journal} {\bibinfo  {journal}
  {Nature nanotechnology}\ }\textbf {\bibinfo {volume} {12}},\ \bibinfo {pages}
  {744} (\bibinfo {year} {2017})}\BibitemShut {NoStop}%
\bibitem [{\citenamefont {Hern{\'a}ndez-V{\'a}zquez}\ \emph
  {et~al.}(2022)\citenamefont {Hern{\'a}ndez-V{\'a}zquez}, \citenamefont
  {de~Luna~Bugallo},\ and\ \citenamefont {Olgu{\'\i}n}}]{hernandez2022janus}%
  \BibitemOpen
  \bibfield  {author} {\bibinfo {author} {\bibfnamefont {M.~{\'A}.}\
  \bibnamefont {Hern{\'a}ndez-V{\'a}zquez}}, \bibinfo {author} {\bibfnamefont
  {A.}~\bibnamefont {de~Luna~Bugallo}},\ and\ \bibinfo {author} {\bibfnamefont
  {D.}~\bibnamefont {Olgu{\'\i}n}},\ }\bibfield  {title} {\bibinfo {title}
  {{Janus monolayers of transition metal dichalcogenides: A DFT study}},\
  }\href@noop {} {\bibfield  {journal} {\bibinfo  {journal} {physica status
  solidi (b)}\ }\textbf {\bibinfo {volume} {259}},\ \bibinfo {pages} {2100248}
  (\bibinfo {year} {2022})}\BibitemShut {NoStop}%
\bibitem [{\citenamefont {Chen}\ \emph
  {et~al.}(2022{\natexlab{b}})\citenamefont {Chen}, \citenamefont {Liu},
  \citenamefont {Lu}, \citenamefont {Zhao}, \citenamefont {Hu}, \citenamefont
  {Ren},\ and\ \citenamefont {Yuan}}]{chen2022intrinsic}%
  \BibitemOpen
  \bibfield  {author} {\bibinfo {author} {\bibfnamefont {H.}~\bibnamefont
  {Chen}}, \bibinfo {author} {\bibfnamefont {R.}~\bibnamefont {Liu}}, \bibinfo
  {author} {\bibfnamefont {J.}~\bibnamefont {Lu}}, \bibinfo {author}
  {\bibfnamefont {X.}~\bibnamefont {Zhao}}, \bibinfo {author} {\bibfnamefont
  {G.}~\bibnamefont {Hu}}, \bibinfo {author} {\bibfnamefont {J.}~\bibnamefont
  {Ren}},\ and\ \bibinfo {author} {\bibfnamefont {X.}~\bibnamefont {Yuan}},\
  }\bibfield  {title} {\bibinfo {title} {{Intrinsic valley-polarized quantum
  anomalous Hall effect and controllable topological phase transition in Janus
  Fe2SSe}},\ }\href@noop {} {\bibfield  {journal} {\bibinfo  {journal} {The
  Journal of Physical Chemistry Letters}\ }\textbf {\bibinfo {volume} {13}},\
  \bibinfo {pages} {10297} (\bibinfo {year} {2022}{\natexlab{b}})}\BibitemShut
  {NoStop}%
\end{thebibliography}%

\end{document}